\documentclass[onethmnum]{ws-rv975x65}
\usepackage{subfigure}     % required only when side-by-side / subfigures are used
\usepackage{ws-rv-van}     % numbered citation/references (default)
\usepackage{ws-index}     % to produce multiple indexes

\makeindex
\newcommand{\Sch}{Schr\"odinger }

\def\be{\begin{equation}}
\def\ee{\end{equation}}
\def\Tr{\mathrm{Tr}\,}

\newcommand{\boxedtext}[1]{\medskip\noindent\fbox{\begin{minipage}[h]{4.65in}
#1\end{minipage}}\medskip}

\def\etal{et al.\ }
\def\EJ{E_\mathrm{J}}
\def\EC{E_\mathrm{C}}
\def\EL{E_\mathrm{L}}

\begin{document}

\chapter[S.M. Girvin]{Basic Concepts in Quantum Information}\label{ra_ch1}

\author[S. M. Girvin]{Steven M. Girvin}%%%%%%%%%%%%%%%\footnote{Author footnote.}}
%\index[aindx]{Author, F.} % or \aindx{Author, F.}
%\index[aindx]{Author, S.} % or \aindx{Author, S.}

\address{Sloane Physics Laboratory\\
Yale University\\
New Haven, CT 06520-8120 USA \\
steven.girvin@yale.edu}%%%%%%%%%%%%%\footnote{Affiliation footnote.}}

\begin{abstract}
In the last 25 years a new understanding has evolved of the role of information in quantum mechanics.  At the same time there has been tremendous progress in atomic/optical physics and condensed matter physics, and particularly at the interface between these two formerly distinct fields, in developing experimental systems whose quantum states are long-lived and which can be engineered to perform quantum information processing tasks.  These lecture notes will present a brief introduction to the basic theoretical concepts behind this `second quantum revolution.'  These notes will also provide an introduction to `circuit QED.' In addition to being a novel test bed for non-linear quantum optics of microwave electrical circuits which use superconducting qubits as artificial atoms, circuit QED offers an architecture for constructing quantum information processors.
\end{abstract}

\body

\section{INTRODUCTION and OVERVIEW}

Quantum mechanics is now more than 85 years old and is one of the most successful theories in all of physics.  With this theory we are able to make remarkably precise predictions of the outcome of experiments in the microscopic world of atoms, molecules, photons, and elementary particles.  Yet despite this success, the subject remains shrouded in mystery because quantum phenomena are highly counter-intuitive to human beings whose intuition is built upon experience in the macroscopic world.  Indeed, the founders of the field struggled mightily to comprehend it.  Albert Einstein, who made truly fundamental contributions to the theory, never fully believed it.  Ironically this was because he had the deepest understanding of the counter-intuitive predictions of the theory.  In his remarkable `EPR' paper with Podolsky and Rosen\cite{EPR}, he famously proposed a thought experiment showing that quantum entanglement produced experimental outcomes that seemed to be obviously impossible.  In an even greater irony, modern experiments based on John Bell's analysis of the EPR `paradox' now provide the best proof not only that quantum mechanics is correct, but that reality itself is simply not what we naively think it is.
In the last 25 years we have suddenly and quite unexpectedly come to a deeper understanding of the role of information in the quantum world.    These ideas could very easily have been understood in 1930 because of the laws of quantum mechanics were known at the time.  Yet, because of the counter-intuitive nature of these concepts, it would be nearly another 60 years before these further implications of the quantum laws came to be understood.  We begin with a high-level `executive summary' overview of these concepts and ideas which will then be developed in more detail in later sections.

From the earliest days of the quantum theory, the Heisenberg uncertainty principle taught us that certain physical quantities could never be known to arbitrary precision.  This, combined with the genuine randomness of the outcome of certain measurements, seemed to indicate that somehow we were `worse off' than we had been in the old world of classical physics.  Recently however, we have come to understand that quantum uncertainty, far from being a `bug', is actually a `feature' in the program of the universe.  Quantum uncertainty and its kin, entanglement, are actually powerful resources which we can use to great advantage to, encode, decode, transmit and process information in highly efficient ways that are impossible in the classical world.   In addition to being a profound advance in our understanding of quantum mechanics, these revolutionary ideas have spurred development of a new kind of quantum engineering whose goal is to develop working quantum `machines.'\cite{LesHouchesQuantumMachines}  The new ideas that have been developed also have profound applications for precision measurements and atomic clocks.  For example, ordinary atomic clocks keep track of time by following the evolution of product states of $N$ spins, each precessing at frequency $\Omega_0$:
\begin{equation}
|\Psi(t)\rangle = [|0\rangle + e^{-i\Omega_0t}|1\rangle]\otimes [|0\rangle + e^{-i\Omega_0t}|1\rangle]\otimes\ldots\otimes [|0\rangle + e^{-i\Omega_0t}|1\rangle].
\end{equation}
By using maximally entangled states of the form
\begin{equation}
|\Psi(t)\rangle = [|0000\ldots 0\rangle + e^{-iN\Omega_0t}|1111\ldots 1\rangle]
\end{equation}
one gets a clock which runs $N$ times faster and therefore gains a factor of $\sqrt{N}$ in precision.\cite{SpinSqueezing1,SpinSqueezing2}

\subsection{CLASSICAL INFORMATION AND COMPUTATION}

For a discussion of the history of ideas in classical and quantum computation, the reader is directed to the central reference work in the field `Quantum Computation and Quantum Information,' by Nielsen and Chuang \cite{MikeandIke}.  The brief historical discussion in this and the following section relies heavily on this work.

The theory of computation developed beginning in the 1930's by Turing, Church and others, taught us that one can, in principle, construct universal computational engines which can compute anything that can be computed.  (We also learned that not everything is computable.)  Any given `Turing machine' can be programmed to perfectly emulate the operation of any other computer and compile any other programming language.  Thus one of the founding principles of computer science is that the actual hardware used in a computation is (formally at least) irrelevant.

A key measure of the difficulty of a computational problem is how the computational cost scales with the size N of the input data set.  Skipping over many essential details, the main classification distinction is between polynomial and exponential scaling.  For example, solving the \Sch equation for N electrons (generically) becomes exponentially harder as N increases.  So does the classical problem of the traveling salesman who has to find an optimal route to visit N cities. Turing's ideas taught us that if a problem is exponential on one piece of hardware, it will be exponential on any piece of hardware.  All universal computers are `polynomially equivalent.'  That is, the scaling is the same up to possible factors no stronger than polynomial in N.
It came as a great shock to the world of physicists and computer scientists that if one could build a computer based on the principles of quantum mechanics, certain classes of problems would change from exponential to polynomial.  That is, not all hardware is equivalent.   Peter Shor \cite{Shor1994} developed a remarkable stochastic quantum algorithm for finding the prime factors of large integers which scales only as $(\ln N)^3$,  an enormous exponential speed up over the fastest known classical algorithm.  [We should be careful to note here that, despite vast effort, it has not been rigorously proven that this particular example (factoring) is an exponentially hard problem classically.]

There are several threads leading to the conclusion that quantum hardware could be very powerful.  One thread goes back to Feynman \cite{Feynman1982} who argued in the 1980's that since the \Sch equation is hard to solve, perhaps small quantum systems could be used to simulate other quantum systems of interest.  It is felt in some quarters that Feynman tends to be given a bit too much credit since this idea did not directly lead to the greater understanding of information and computation in quantum mechanics.  However he definitely did pose the question of whether quantum hardware is different than classical.  There are active attempts underway today to build `quantum simulators' using optical lattices, trapped ions and other quantum hardware.  This effort flows both from Feynman's ideas and from the new ideas of quantum information processing which will be described below.

\subsection{Quantum Information}
A classical bit is some physical system that has two stable states, 0 and 1.  Quantum bits are quantum systems that have two energy eigenstates $|0\rangle$ and $|1\rangle$.  They derive their extraordinary power from being able to be in coherent quantum superpositions of both states.  This `quantum parallelism' allows them (in some sense) to be both 0 and 1 at the same time.

One additional point worth discussing before we proceed is the following.  Classical computers dissipate enormous amounts of energy.  (The fraction of electrical power consumed in the U.S. by computers is of order 3\% of the total electrical consumption from all sectors of the economy.)  Quantum computers work using coherent quantum superpositions and hence fail if {\em any} energy is dissipated during the computation.  That is, the unitary time evolution of a quantum computer is fully reversible.  (I ignore here quantum error correction, described below, in which a small amount of energy must be dissipated to correct imperfections in the hardware operation.)  In thinking about the thermodynamics of information, Charles H. Bennett \cite{BennettReversible} made a major contribution in 1973 by discovering the concept of reversible classical computation.  He proved that energy dissipation is not inevitable and thus fully reversible computation is possible in principle.   It can be argued that this unexpected idea helped lay the ground work for quantum information processing which, as noted, must of necessity be fully reversible.

The first hint that quantum uncertainty might be a `feature' and not a `bug,' came from a remarkably prescient idea of Stephen Wiesner, then a graduate student at Columbia, sometime around 1968 or 1970.  The idea was so revolutionary that Wiesner was unable to get it published until 1983\cite{Wiesner1983}. The idea was that one could create `quantum money'\cite{AaronsonQuantumMoney} with quantum bits for the serial number and this would make it possible to verify the validity of the serial number, but never duplicate  it to make counterfeit money.  This was really the first understanding of the `no cloning theorem' later formally proved by Wooters and Zurek \cite{Wootters1982} and Dieks \cite{Dieks1982}.   The essential idea is that if we measure the polarization of a two-level system (a quantum bit, say a spin-1/2 particle), there are always only two possible answers, $\pm 1$, no matter what the state of the qubit is.  Thus, the result of our measurement yields only one classical bit of information.  However unless you know what quantization axis to choose for the measurement, you will most likely destroy the original state during the measurement process.   This is because the result of any measurement of the polarization of the spin along some axis always shows that the qubit is maximally polarized with projection $\pm 1$ along the axis you chose to measure (as opposed to the axis along which the state was prepared).  The act of measurement itself collapses the state and so no further information can be acquired about the original state.  Since it takes an infinite number of classical bits to specify the original orientation of the quantization axis, we cannot learn the quantum state from the one bit of classical information produced by the measurement.    On the other hand, if you are told what axis to use in the measurement, then you can determine whether the polarization is $+1$ or $-1$ along that axis with complete certainty and without destroying the state, since the measurement is QND (quantum non-demolition).

Wiesner realized that all this means that one could make a quantum serial number by randomly choosing the orientation of a spin for each of, say $N=20$ qubits.  For each qubit the quantization axis could be randomly chosen to be X or Z and the sign of the polarization along that axis could be chosen randomly to be $\pm 1$.  The agency that originally created the quantum serial number could verify the validity of the money by making QND measurements with the correct quantization axes (which having chosen the quantization axes, it alone knows) thus obtaining the correct polarizations (if the serial number is not counterfeit), but a counterfeiter (who does not have access to the information on the randomly chosen quantization directions) has a low probability (which turns out to be $(3/4)^N$) of being able to make a copy of the serial number that would pass inspection by the agency.  This powerful result is a simple consequence of the Heisenberg uncertainty principle and the fact that the $X$ and $Z$ components of the spin are incompatible observables.  As such it could have been `obvious' to practitioners of quantum mechanics anytime after about 1930, but it was not understood until 40 years later, and then by only one person.

Though Wiesner could not initially get his idea published, one person who paid attention was Charles Bennett \cite{Bennett_privcomm,Divincenzo_privcomm}, now at IBM Corporation.  Bennett, working with Gilles Brassard, realized in 1984 that the quantum money idea could be used create an unbreakable quantum encryption protocol.  The only provably secure encryption technique uses a secret key called a `one-time pad.'  The key is a random string of bits as long as the message to be sent and is used to code and decode the  message.  The problem is that the sender and receiver must both have a copy of the key and it can only be used once.  The difficulty of how to secretly transmit the key to the receiver is identical to the original problem of secretly transmitting the message to the receiver and is called the `key distribution problem.'  Bennett and Brassard's so-called BB84 protocol \cite{BB84} for quantum key distribution solves the problem by transmitting the key as a string of quantum bits (essentially as the serial number of a piece of quantum money).  Because of the no-cloning theorem, it is not possible for an eavesdropper to read the key and pass it on undisturbed in order to avoid detection.  From this fact, BB84 provides a protocol for insuring that no eavesdropper has read the key and thus that it is safe to use.  The BB84 protocol was a major breakthrough in our understanding of quantum information and its transmission.  It too was sufficiently revolutionary that the authors had difficulty getting it published \cite{Divincenzo_privcomm}.  The BB84 protocol and its descendants have led to practical applications in which quantum key distribution can now be routinely done over long distances at megahertz bit rates using the polarization modes of individual photons as the quantum bits.  It has been carried out over optical fiber and through the atmosphere over distances of many tens of kilometers and there are now available commercial systems based on these ideas.  Quantum cryptography is reviewed by Gisin et al.\cite{GisinRMP2002}.

Following up on the idea of quantum key distribution, Bennett and Wiesner invented the idea of `quantum dense coding' \cite{quantumdensecoding}.  Given that measuring a qubit yields only one classical bit of information, it is not possible to transmit more information with a single quantum bit than with a single classical bit.  However if the sender and receiver each share one quantum bit from an entangled pair, something remarkable can occur.  There are four standard `Bell states' for an entangled pair ($|01\rangle\pm|10\rangle, |00\rangle \pm |11\rangle$).  Suppose that Alice prepares the Bell state $|00\rangle + |11\rangle$, and then sends one of the two qubits to Bob.  So far, no information has been transmitted, because if Bob makes a measurement on his qubit, the result is simply random, no matter what axis he chooses for his measurement.  However (as we show in detail in a later section), Bob can now perform one of four local unitary operations on his qubit to either leave the state alone or map it into one of the three other Bell states.  He then sends the qubit back to Alice who can make a joint measurement \cite{jointmeasurement} on the two qubits and determine which of the four Bell states results.  Since there are four possibilities, Bob has transmitted two classical bits by simply returning one quantum bit to Alice.

Another way to illustrate the peculiar power of entanglement is for Alice to prepare the same Bell state, $|00\rangle + |11\rangle$.  She then transmits the first qubit to Bob. After that she performs one of four possible local unitary operations on her remaining qubit to either leave the state alone or map it into one of the three other Bell states (details will be supplied further below).  Alice then sends this second qubit to Bob who then can measure which of the four states the qubits are in and hence receive two classical bits of information.  This is not surprising since he has received two qubits.  Recall however that Alice did not decide which of the four `messages' to send until {\em after} she had sent the first qubit to Bob.  This result clearly demonstrates the `spooky action at a distance' that bothered Einstein and has further enhanced our understanding of the novel features of information and its transmission through quantum channels.  Following up on the work of Bennett and Brassard, Ekert \cite{EkertQCryptoBell} proposed use of entangled pairs as a resource for quantum encryption, though this is not yet a practical technology.

One of the very important applications of these new ideas in quantum information is to the numerical solution of the \Sch equation for strongly correlated many-body systems.  I. Cirac, G. Vidal, F. Verstrate, and others have realized that quantum information ideas are very helpful in understanding how to efficiently represent in a classical computer the highly entangled quantum states that occur in strongly correlated many-body systems.  This has led to real breakthroughs in our understanding and modeling of complex many-body systems\cite{many-body-quantum-info}.

\subsection{Quantum Algorithms}

As Feynman noted, simulating quantum systems on a classical computer is exponentially hard.  It is clear that a quantum computer can efficiently simulate certain quantum systems (e.g. the computer can at least simulate itself, and a piece of iron can be used to simulate iron!)
The next development in going from quantum information to quantum computation was taken by David Deutsch who posed the question of whether there is a quantum extension of the Church-Turing idea that any computation running on a classical computer can be {\em efficiently} simulated on a universal Turing machine.  Deutsch asked if there exists a universal quantum simulator which can efficiently simulate any other quantum system.  The answer to this profound question remains unknown, but in 1985 Deutsch \cite{Deutsch1985} opened the door to the world of quantum algorithms.  He found a `toy' computational problem that could be solved on a quantum computer in a manner that is impossible classically.   In 1992 Deutsch and Jozsa \cite{Deutsch-Jozsa1992} simplified and extended the earlier result.

The modern formulation of the problem is the following.  Consider a function $f(x)$ whose domain is $x=\pm1$ and whose range is also $\pm 1$.  There are exactly four such functions:  $f_1(x) = x, f_2(x) = -x, f_3(x) = +1, f_4(x) = -1$.  If we measure $f(-1)$ and $f(+1)$ we acquire two bits of classical information and know which of the four functions we have.  Obviously this requires two evaluations of the function.  Now, notice that the functions can be divided into two classes, balanced (i.e. $f(-1)+f(+1)=0$) and constant (i.e. unbalanced).  Since there are only two classes, finding out which class the function is in means acquiring one classical bit of information.  However, it still requires two evaluations of the function. Hence classically, we are required to learn the full function (by making two evaluations) before we can determine which class it is in.  Quantum mechanically we can determine the class with only a single measurement!  The trick is to use the power of quantum superpositions.  By putting $x$ into a superposition of both $+1$ and $-1$, a single evaluation of the function can be used to determine the class (but {\em not} which function in the class since the measurement yields only one bit of classical information).

This idea launched searches for other quantum algorithms.  Consider the problem of finding a particular entry in a large unordered database.  (For example imagine looking for a friend's phone number in a phone book that was not alphabetically ordered).  Classically there is no faster procedure than simply starting at the beginning and examining the database entries one at a time until the desired entry is found.  On average this requires $N/2$ `looks' at the database.  Lov Grover \cite{Grover1996,Grover1997} found a quantum search algorithm that allows one to find an entry in an unordered database in only $\sim \sqrt{N}$ `looks,' which is a speed up by a factor $\sim\sqrt{N}$ over the classical search.  Again, the superposition and uncertainty principles come into play.  A given `look' can actually be a superposition of single looks at every possible entry at the same time.  Since searching is a very generic problem in computer science, this quantum result is quite significant, even though the speedup is not exponential.

The real excitement came in 1994 when Peter Shor \cite{Shor1994,Shor1997} found efficient quantum algorithms for the discrete logarithm problem and the Fourier transform problem.  The latter can be used to find the prime factors of large integers and hence break RSA public key encryption.  The classical Fourier transform is defined by the linear transformation
\begin{equation}
\tilde f(k) = \frac{1}{\sqrt{N}}\sum_{j=0}^{N-1} e^{2\pi ikj/N} f(j)
\end{equation}
and requires of order $N^2$ operations to evaluate the function $\tilde f$.   For the special case that $N=2^n$, the Fast Fourier Transform (FFT)
algorithm requires only $\sim N\ln N = n2^n$ steps, but is still exponential in $n$.  Shor's quantum Fourier transform is a unitary transformation in Hilbert space which acts on a state defined by the function $f$ and produces a new state defined by the function $\tilde f$.  Suppose we have $n$ qubits so that the Hilbert space has dimension $2^n$.  We can use $f$ to define the following (unnormalized) state
\begin{equation}
|\Psi\rangle = \sum_{j=0}^{2^n-1} f(j)|j\rangle.
\end{equation}
The quantum Fourier transform applied to this state is defined by
\begin{equation}
|\tilde\Psi\rangle =U_{\rm QFT}|\Psi\rangle = \sum_{j=0}^{2^n-1} \tilde f(j)|j\rangle.
\end{equation}
Remarkably, this operation can be carried out in order $n^2$ steps (more precisely order $n^2 \ln n \ln\ln n$).  This is an {\em exponential speed up} over the classical FFT procedure.

The bad news is that being in possession of the quantum state $|\tilde\Psi\rangle$ does not actually tell you the values of $\tilde f(j)$.  Determining these would in general require exponentially many measurements. This is because a measurement of which state the qubits are in randomly yields $|j\rangle$ with probability $S(j)\equiv |\tilde f(j)|^2$. Suppose however that $f$ is periodic with an integer period and is simple enough that its spectral density $S(j)$ is strongly peaked at some  value of $j$.  Then with only a few measurements one would find $j$ with high probability and hence the period could be found.  Peter Shor mapped the prime factor problem onto the problem of finding the period of a certain function and hence made it soluble.  This result caused a sensation and thereafter the field of quantum information and computation exploded with feverish activity which continues unabated.

With this overview, we are now ready to begin a more detailed discussion.  Two interesting  topics which will not be covered in this discussion are quantum walks\cite{Kempe_quantum_walks} and adiabatic quantum computation.\cite{Farhi_Adiabatic_2001,Farhi_Adiabatic_2012}

\section{Introduction to Quantum Information}\label{ra_sec1}

In recent decades there has been a `second quantum revolution' in which we have come to much (but still not completely) understand the role of information in quantum mechanics.  These notes provides only the briefest introduction to a few of the most basic concepts.  For fuller discussions the reader is directed to Nielsen and Chuang\cite{MikeandIke} and Mermin\cite{MerminBook}.

The bit is the smallest unit of classical information.  It represents the information we gain when we receive the definitive answer to a yes/no or true/false question about which we had no prior knowledge.   A bit of information, represented mathematically as binary digit 0 or 1, can be represented physically inside a computer in the form of two physical states of a switch (on/off) or the voltage state of some transistor circuit (high/low).  One of the powerful features of this digital (as opposed to analog) encoding of information is that it is robust against the presence of noise as long as the noise is small compared to the difference in signal strength between the high and low states.

It is useful to note that there are two possible encodings of a mathematical bit in a physical bit.  The low voltage state can represent 0 and the high voltage state can represent 1, or we can use the reverse.  It does not matter which encoding we use as long as everyone using the information agrees on it.   If two parties are using opposite encodings, they can still readily translate from one to the other by performing a NOT operation (which maps 0 to 1 and vice versa) to each bit.

The situation is very different for quantum bits (qubits).  A quantum bit is represented physically by an atom or other quantum system which has a two-state Hilbert space.  For example the state (in Dirac notation) $|0\rangle$ could be represented physically by the atom being in the ground state and the state $|1\rangle$ could be represented by the atom being in its first excited state. (We assume that the other excited states can be ignored.)  All two-state quantum systems are mathematically equivalent to a spin-1/2 particle in a magnetic field with Hamiltonian
\be
H=\frac{\hbar}{2}\vec \omega\cdot \sigma,
\ee
where $\vec\omega$ is a vector (with units of frequency) representing the strength and direction of the pseudo magnetic field, and $\vec \sigma = (\sigma^x,\sigma^y,\sigma^z)$ are the Pauli spin matrices.  The transition frequency from ground to excited state is $\omega \equiv |\vec\omega|$.
Up to an additive constant, this form of Hamiltonian is the most general Hermitian operator on the two-dimensional Hilbert space.  That is, the Pauli operators $(\sigma^x,\sigma^y,\sigma^z)$ for the three components of the `spin' form a basis which spans the set of all possible operators (ignoring the identity operator).
Because they are non-commuting, they are mutually incompatible and it is not possible to have a simultaneous eigenstate of more than one of the operators.  The most general measurement that can be made is a `Stern-Gerlach' measurement of the projection of the spin along a single axis
\be
{\cal O}={\hat n} \cdot\vec\sigma.
\ee
Since ${\cal O}^2=1$, the eigenvalues of ${\cal O}$ are $\pm1$ and such a Stern-Gerlach measurement yields precisely one classical bit of information telling us whether the spin is aligned or anti-aligned to the axis $\hat n$.  It is a strange feature of quantum spins that no matter what axis we choose to measure the spin along, we always find that it is precisely parallel or antiparallel to that axis.

The ideal Stern-Gerlach measurement is quantum non-demolition (QND), that is it is repeatable.  If we measure (say) $\sigma^z$ for an unknown quantum state, the result will be (possibly randomly) $\pm1$. As long as there are no stray magnetic fields which cause the spin to precess in between measurements\footnote{That is, a measurement is QND if the operator being measured and its coupling to the measurement apparatus both commute with the system Hamiltonian.}, subsequent measurements of $\sigma^z$ will not be random but will instead be identical to the first.  If this is followed by a measurement of (say) $\sigma^x$ the result will be completely random and unpredictable with $+1$ and $-1$ each occuring half the time.  However subsequent measurements of $\sigma^x$ will all agree with the first if the measurement is QND.

The spinor eigenfunctions of $H$ are (up to an unimportant global phase factor)
\begin{eqnarray}
|\psi_+\rangle &=& \left(\begin{array}{c}\cos\frac{\theta}{2}e^{i\varphi/2}\\
\sin\frac{\theta}{2}e^{-i\varphi/2}\end{array}\right),\\
|\psi_-\rangle &=& \left(\begin{array}{c}-\sin\frac{\theta}{2}e^{i\varphi/2}\\
\cos\frac{\theta}{2}e^{-i\varphi/2}\end{array}\right),
\end{eqnarray}
where $\theta$ and $\varphi$ are respectively the polar and azimuthal angles defining the direction of $\vec\omega$.
For the case $\theta=\varphi=0$, the Hamiltonian reduces to
\be
H_0 = \frac{\omega}{2}\sigma^z,
\ee
with eigenfunctions that are simply the `spin up' and `spin down' basis states
\begin{eqnarray}
|0\rangle&=&|\downarrow\rangle=\left(\begin{array}{c} 0\\1\end{array}\right)\\
|1\rangle &=& |\uparrow\rangle=\left(\begin{array}{c} 1\\0\end{array}\right).
\end{eqnarray}
Throughout these notes we will use the $0,1$ and the $\downarrow,\uparrow$ notation interchangeably.

We see from the above that, ignoring any overall phase factor, it takes two real numbers (or equivalently an infinite number of classical bits) to specify the quantum state of a spin. The huge asymmetry between the infinite number of classical bits needed to specify a state and the single classical bit we can obtain by doing a measurement is a key concept in our understanding of quantum information, quantum encryption and quantum information processing.   The act of measurement collapses the spin state onto the measurement axis and we know exactly what the state of the spin is after the measurement.  However (unless we were told in advance which axis to use) we have no way of knowing what the actual state of the spin was before the measurement.  This leads us directly to the no-cloning theorem\cite{Wootters1982} which states that it is impossible make a copy of an unknown quantum state.  If we are given an unknown state the only way we can copy it is to make a measurement to see what the state is and then orient additional spins to match that state.  However since the state is unknown and we are only allowed to make one measurement, we are forced to guess which axis to measure along and we cannot guarantee that the state has not been changed by the act of measurement.

The mathematical version of the argument for the no-cloning theorem is the following.  Let us start with a product state of a qubit in an unknown superposition state and an ancilla qubit in the $|0\rangle$ state
\be
|\Psi\rangle = [\alpha|0\rangle + \beta|1\rangle]\otimes|0\rangle.
\ee
After the cloning operation we desire to have both qubits in the same state
\begin{eqnarray}
|\Psi_\mathrm{clone}\rangle &=& [\alpha|0\rangle + \beta|1\rangle]\otimes[\alpha|0\rangle + \beta|1\rangle]\\
|\Psi_\mathrm{clone}\rangle &=& \alpha^2|00\rangle + \alpha\beta|10\rangle +\alpha\beta|01\rangle + \beta^2|11\rangle].
\end{eqnarray}
The only operations which are physically possible in quantum mechanics are represented by unitary transformations.
Thus we seek a transformation $U$ obeying
\be
|\Psi_\mathrm{clone}\rangle = U|\Psi\rangle.
\ee
  U is a linear transformation and so it is impossible for it to produce the non-linear coefficients $\alpha^2,\alpha\beta,\beta^2$ that we see above.  Hence cloning an unknown state is impossible.  Of course if the state is known, we could use a unitary transformation $U(\alpha,\beta)$ that explicitly depends on the known parameters $\alpha$ and $\beta$ and successfully achieve cloning.  We cannot do so however with a unitary transformation which does not depend on knowledge of $\alpha$ and $\beta$.  Thus known states are readily cloned, but unknown states cannot be.  This result will have profound implications for the encryption of information using quantum states and for quantum error correction.

We can understand the no cloning theorem from another point of view.  Quantum bits also encode only a single classical bit (the measurement result), but unlike the classical case where there are only two possible encoding schemes, there are an infinity of different quantum encodings defined by the axis $\hat n = (\sin\theta\cos\varphi,\sin\theta\sin\varphi,\cos\theta)$ that was used for the preparation and should be used for the measurement if the information is to be correctly decoded.  For the classical case, the transformation from one encoding scheme to the other is simply the NOT operation $0\rightarrow 1, 1\rightarrow 0$.  For the quantum case the transformation is a unitary operation which rotates the spin states on the Bloch sphere through the appropriate angle connecting the old axis $\hat n$ and the new axis $\hat n^\prime$
\be
U_{\hat n^\prime \hat n} = e^{i\frac{\Theta}{2}\hat v\cdot\vec\sigma},
\ee
where $\cos\Theta=\hat n^\prime\cdot\hat n$ defines the angle between the two axes and $\hat v = \frac{\hat n^\prime \times \hat n}{|\hat n^\prime \times \hat n|}$ is the unit vector perpendicular to the plane formed by the two axes.  If we are not given any information about which of the infinitely many encoding schemes have been used we cannot reliably decode the information.  If we do know the encoding axis $\hat n$, the above transformation law will tell us what the results will be for measurements done in a different `decoding' basis $\hat n^\prime$.  In particular it tells us that if we encode on the z axis and decode on the x axis the results of the decoding will be totally random and uncorrelated with the encoding.  For example, we can express $|\uparrow\rangle, |\downarrow\rangle$, the $\pm 1$ eigenstates of $\sigma^z$ as a superposition of the eigenstates of $\sigma^x$
\begin{eqnarray}
|\uparrow\rangle &=& \frac{1}{2}[|\rightarrow\rangle + |\leftarrow\rangle]\\
|\downarrow\rangle &=& \frac{1}{2}[|\rightarrow\rangle - |\leftarrow\rangle],
\end{eqnarray}
which confirms the statement made above that if we make a $\sigma^x$ measurement after a $\sigma^z$ measurement, the results will be $\pm 1$ with equal probability.  This is distinctly different than the classical case where, if we accidentally use the wrong decoding scheme, the results are still deterministic and perfectly correlated with the encoded data (just backwards with 0 and 1 being interchanged).

Further below, we will discuss quantum teleportation.  This is a process in which an unknown quantum state is perfectly reproduced at a distant location.  This is not restricted by the no-cloning theorem because the original state is destroyed during the process of teleportation.  For example teleportation could consist of simply swapping two states
\begin{eqnarray}
|\Psi\rangle &=& [\alpha|0\rangle + \beta|1\rangle]\otimes|0\rangle,\\
|\Psi_\mathrm{teleport}\rangle &=& |0\rangle \otimes [\alpha|0\rangle + \beta|1\rangle].
\end{eqnarray}
Because the coefficients $\alpha$ and $\beta$ enter only linearly, there exists a unitary transformation which can accomplish this task.  We will explore a particular protocol for how this can be achieved using entangled states further below.

  \section{Quantum Money and Quantum Encryption}

  In the 1970's Steven Wiesner, then a graduate student at Columbia, came up with the idea of `quantum money' which cannot be counterfeited.   This was the very first idea in a chain of ideas that has led to the revolution in quantum information.  Legend has it that the idea was so far ahead of its time that Wiesner was unable to get it published until much later\cite{Wiesner1983}.  He did however discuss it with Charles Bennett and a decade later it bore fruit in the form of quantum encryption, about which more below.

  The idea of quantum money is essentially based on the no-cloning theorem, although that theorem was not formally stated and proved until sometime later\cite{Wootters1982}.  The idea is to create a serial number using quantum bits which encode a random string of 0's and 1's.  The quantum wrinkle is that for each bit, one randomly selects the $Z$ encoding ($\hat n = \hat z$) or the $X$ encoding ($\hat n = \hat x$).  Thus the serial number might look like this:  $|\uparrow,\rightarrow,\rightarrow,\downarrow,\leftarrow,\leftarrow,\ldots\rangle$.  Because of the no-cloning theorem, it is impossible for a counterfeiter to make copies of the money and its quantum serial number.  The counterfeiter would have to guess which measurement to make ($X$ or $Z$) on each qubit and would have no way of knowing if she had guessed correctly.   On the other hand, the treasury department could easily verify whether any given bill is real or counterfeit.  If the bill also carries an ordinary classical serial number that uniquely identifies it, then the treasury department can keep a (secret) record of the bit value and the quantization (encoding) axis used for each qubit on that particular bill (labeled by the classical serial number) and so knows which measurement to make for each one.  If for example the first bit is in state $|\uparrow\rangle$, a $Z$ measurement will always yield the correct value of $+1$ with no randomness.  If the treasury department uses all the correct measurement orientations and does not recover the correct measurements then the bill is counterfeit (or the qubits have decohered!).

  \boxedtext{
  \begin{problem}
  Prove that the probability of a counterfeit `copy' of a bill with an $N$ qubit quantum serial number succesfully passing the treasury department's scrutiny is $\left(\frac{3}{4}\right)^N$.  Thus for large $N$, counterfeiting is very unlikely to succeed.  Hint: in order for a particular bit to fail the test, the counterfeiter must have chosen the wrong measurement and the treasury has to be unlucky in its measurement result.

  Show that this result remains valid even if the counterfeiter chooses an arbitrary pair of orthogonal axes, $X',Z'$ to make her measurements, as long as they lie in the same plane as the $X,Z$ axes chosen by Bob and Alice.
  \end{problem}
  }

  \subsection{Classical and Quantum Encryption}
  The only provably secure method of classical encryption is the `one-time pad.'  Suppose that Alice has a message which consists of a string of $N$ classical bits.  Alice wishes to send this message to Bob in such a way that Eve cannot eavesdrop on the communication by intercepting and deciphering the message.  A one-time pad is a string of random bits (also called the encryption `key') whose length is at least as large as the message.  If Bob and Alice each are in possession of identical copies of the one-time pad then the message can be sent via the following simple protocol.  Let the $j$th bit in Alice's message be $M_j$ and the $j$th bit in the one-time pad be $P_j$.  Then Alice computes the $j$th bit in the encrypted message via
  \be
  E_j = M_j \oplus P_j,
  \ee
  where $\oplus$ means addition modulo 2.  Essentially this encryption rule means that if $P_j=0$, do nothing to the message bit and if $P_j=1$, then flip the message bit. Because $E_j$ is completely random and uncorrelated with any of the other encrypted bits, Eve is unable to decipher the message even if she intercepts it during the transmission to Bob.  Bob however is able to decrypt it by the same operation using the same key (one-time pad)
  \be
  D_j = E_j \oplus P_j.
  \ee
  Because $P_j\oplus P_j=0$, we have $D_j = M_j$ and Bob recovers the original message.

  In order to work, the key must be as long as the original message, must be completely random, must be kept secret, and must never be re-used.  While perfectly secure, this method suffers from the so-called `key distribution problem.'  Securely transmitting a one-time pad of length $N$ from Alice to Bob has the same difficulty as the problem of securely transmitting the original message!  Bennett and Brassard\cite{BB84} realized that quantum mechanics provides a solution to this problem.  One can simply distribute the one-time pad as the serial numbers of quantum money!

\begin{figure}[ht]
\centerline{\psfig{file=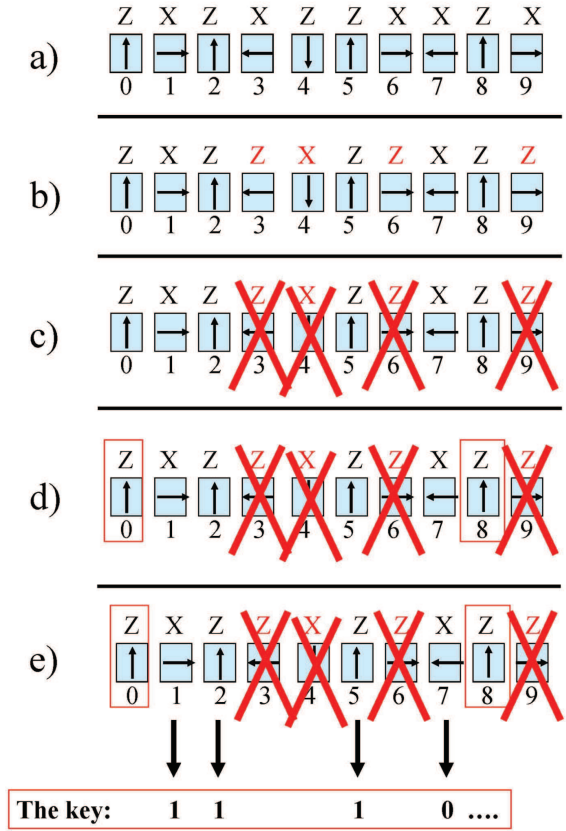,width=3.0in}}
\caption{BB84 protocol for quantum key distribution. a) Alice sends Bob a series of quantum bits chosen randomly from among four (non-orthogonal) states: $+Z,-Z,+X,-X$. b) For each bit Bob randomly guesses whether to measure $Z$ or $X$. Approximately half of his guesses are correct. c) Bob publicly announces which measurements he chose but not the results.  Alice tells him which ones to throw away because he guessed incorrectly. d) Bob publicly announces the measurement results from a randomly selected subset of $M$ of the remaining bits (indicated by the extra surrounding boxes).  If Alice confirms that all the results are correct, then there is a very low probability, $P_M=(3/4)^M$ that an eavesdropper has intercepted the communication. e) Alice and Bob know that Bob's remaining bit measurements agree perfectly with what Alice sent and these are used to create the key using the mapping:  $+Z\rightarrow 1, -Z\rightarrow 0, +X\rightarrow 1, -X\rightarrow 0$.}
\label{fig:quantumkeydistribution}
\end{figure}

  The so-called BB84 protocol developed by Brassard and Bennett illustrated in Fig.~(\ref{fig:quantumkeydistribution}) works as follows.  Alice sends a long string of quantum bits to Bob which have been encoded using randomly selected $Z$ and $X$ encoding (just as for quantum money).  Bob does not know what the correct measurements are to make so he guesses randomly.  He then announces publicly what measurement axis he chose for each qubit but keeps secret the measurement results.  Typically he will have guessed correctly about half the time.  Alice then publicly states which measurements Bob performed correctly and Bob discards the others.  He now has about half the number of qubits as before but he knows that he made the correct measurements on them.  Thus he should have the correct bit values for the key.  But, how can he be sure that Eve has not intercepted the message?  If she has, she will necessarily corrupt some of the results before passing them along to Bob (which she must do in order to conceal her presence).  In essence what she is passing along is a counterfeit bill.  How can Bob check that this has or has not occurred?  The next step in the protocol is that Bob selects a subset of size $M$ of the (good) qubit measurement results at random and publicly announces them.  Alice then publicly announces whether Bob got them right.  If he got them right then the probably that Eve was intercepting the message before passing it on is very low, $P_M=\left(\frac{3}{4}\right)^M$ and it is safe to use the remaining results (which are known to both Bob and Alice) as the encryption key.
If the public results do not match, then Eve's presence has been detected and Alice and Bob do not proceed with using the corrupted key.  (If they did proceed, neither Eve nor Bob would be able to read the message because the cloned key is necessarily corrupted with high probability.)

\section{Entanglement, Bell States and Superdense Coding and Grover Search and Teleportation}

%\boxedtext{
%Discuss lack of information transfer in Bell collapse but double info transfer if one classical bit is subsequently sent.  This unifies the EPR %mystery and quantum dense coding!}

So far we have discussed only the states of a single qubit.  As we shall see, multi-qubits states will prove to be remarkably rich.
A classical register holding $N$ bits, can be in any one of $2^N$ states.  These states can represent, for example, all $N$-bit binary numbers from $0$ to $2^N-1$.  For the quantum case we have studied single qubits which have a Hilbert space spanned by 2 orthogonal states.  For $N$ qubits, we have a Hilbert space which is exponentially large, spanned by $2^N$ orthogonal basis states.  We can label these basis states with binary numbers just as for the classical case.  However in the quantum case, the Hilbert space includes states which are arbitrary superpositions of all the basis states.  While the coefficients for the amplitudes entering such superpositions are continuous complex numbers, let us, for simplicity of counting, restrict our attention to coefficients that are $\pm 1$ so that
\begin{eqnarray}
|\Psi\rangle = \frac{1}{\sqrt{2^N}}\big\{&\pm&|000\ldots 000\rangle \pm |000\ldots 001\rangle \pm |000\ldots 010\rangle\pm |000\ldots 011\rangle\nonumber \ldots\\
&\pm&|111\ldots100\rangle \pm |111\dots 101\rangle\pm|111\dots 110\rangle\pm|111\dots 111\rangle  \big\}
\end{eqnarray}
Because we have $2^N$ sign choices (ignoring the fact that the overall sign is irrelevant) the total number of states is enormously large: $2^{(2^N)}$.  Thus even using only this highly restricted set of states, the storage capacity of quantum memories is truly enormous.  While this capacity might be useful at intermediate steps in a quantum computation, let us never forget that when we readout the memory we always obtain only $N$ classical bits of information representing which of the $2^N$ measurement basis states the memory collapses onto.

Let us focus our attention for the moment on the case of $N=2$ for which there are four orthogonal basis states.  A \emph{separable state} is one which can be written as the product of a state for the first qubit and a state for the second qubit
\begin{eqnarray}
|\Psi\rangle &=& [\alpha_1|0\rangle + \beta_1|1\rangle]\otimes[\alpha_2|0\rangle + \beta_2|1\rangle]\nonumber\\
&=& \alpha_1\alpha_2|00\rangle + \alpha_1\beta_2|01\rangle+\beta_1\alpha_2|10\rangle+\beta_1\beta_2|11\rangle.
\end{eqnarray}
An \emph{entangled state} of two qubits\footnote{The concept of entanglement is more difficult to uniquely define and quantify for $N>2$ qubits.} is any state which cannot be written in this form.  A convenient basis in which to represent entangled states is the so-called Bell basis
\begin{eqnarray}
|B_0\rangle &=&\frac{1}{\sqrt{2}}\left[|\uparrow\downarrow\rangle-|\downarrow\uparrow\rangle\right]\\
|B_1\rangle &=&\frac{1}{\sqrt{2}}\left[|\uparrow\downarrow\rangle+|\downarrow\uparrow\rangle\right]\\
|B_2\rangle &=&\frac{1}{\sqrt{2}}\left[-|\uparrow\uparrow\rangle+|\downarrow\downarrow\rangle\right]\\
|B_3\rangle &=&\frac{1}{\sqrt{2}}\left[-|\uparrow\uparrow\rangle-|\downarrow\downarrow\rangle\right].
\end{eqnarray}
Each of these is a maximally entangled state, but they are mutually orthogonal and span the full Hilbert space.  Therefore linear superpositions of them can represent product states.  For example,
\be
|\uparrow\rangle|\uparrow\rangle=-\frac{1}{\sqrt{2}}\left[|B_2\rangle+|B_3\rangle\right].
\ee

Entanglement is very mysterious and entangled states have many peculiar and counter-intuitive properties.  In an entangled state the individual spin components have zero expectation value and yet the spins are strongly correlated.  For example,
\begin{eqnarray}
\langle B_0|\vec\sigma_1|B_0\rangle &=&\vec 0\\
\langle B_0|\vec\sigma_2|B_0\rangle&=&\vec 0\\
 \langle B_0|\sigma_1^x\sigma_2^x|B_0\rangle&=&-1\\
 \langle B_0|\sigma_1^y\sigma_2^y|B_0\rangle&=&-1\\
 \langle B_0|\sigma_1^z\sigma_2^z|B_0\rangle&=&-1.
\end{eqnarray}
This means that the spins have quantum correlations which are stronger than is possible classically.  In particular,
\be
 \langle B_0|\vec\sigma_1\cdot\vec\sigma_2|B_0\rangle=-3
\ee
despite the fact that in any single-spin (or product state) $|\langle \vec\sigma\rangle|\le 1$.

To explore this further, let us consider the concept of \emph{entanglement entropy}\cite{EisertRMP,HorodeckiRMP,many-body-quantum-info,Verstraete}.  In classical statistical mechanics, entropy is a measure of randomness or disorder.  It is proportional to the logarithm of the number of microstates a thermodynamic system can be in given the observed macrostate.   Formally, the entropy of a microstate probability distribution $\{P_k;k=1,N\}$ is given by
\be
S=-k_\mathrm{B}\sum_{j=1}^N P_j\ln P_j,
\ee
(with the convention that $0\ln 0$ is replaced by $\lim_{x\rightarrow 0} x\ln x = 0$) where $k_\mathrm{B}$ is the Boltzmann constant.
In the canonical ensemble of statistical mechanics one assumes that all microstates of a given energy are equally likely.  If there are $N$ such microstates then $P_j=\frac{1}{N}$ and we obtain $S=k_\mathrm{B}\ln N$.  The logarithmic form is important because it means that in thermodynamics, the entropy is an extensive quantity.  Thus for example if we have a system consisting of two subparts with $N_1$ and $N_2$ microstates, the total number of microstates is the product $N_\mathrm{TOT} = N_1N_2$ and the probability of a particular joint microstate is $P_j=\frac{1}{N_1N_2}$, but because of its logarithmic form, the entropy is the sum of the two entropies:  $S=S_1+S_2$.

In information theory, the Shannon entropy of a message drawn from an ensemble of $N$ possible code words is
\be
S=-\sum_{j=1}^N P_j\ln_2 P_j,
\ee
where $P_j$ is the probability that the message will consist of the $j$th codeword and the logarithm is conventionally measured in base 2 so that the entropy (information content) is measured in bits.  For example, if Alice sends Bob a message containing a single physical bit (a 0 or 1), and the a priori probability of the bit being 0 is $P_0=\frac{1}{2}$ and of being 1 is $P_1=\frac{1}{2}$, then $S=1$ bit.  On the other hand if Bob knows ahead of time that Alice almost always sends him a $0$ (because he knows $P_0=0.999$, say), then there will rarely be a surprise in the message and the Shannon information entropy (which is a measure of how much Bob learns on average from the message) is much lower, $S=-0.999\ln_2 0.999-0.001\ln_2 0.001 \approx 0.01$ bits.  If Alice sends Bob two messages (drawn from the same or different ensembles of code words) the logarithmic form guarantees that the total entropy (and hence information content) is additive, $S=S_1+S_2$.
We can connect this to statistical mechanics by noting that if Bob and Alice both look at a \emph{macrostate} of a thermodynamics system and Alice sends Bob the actual \emph{microstate}, the information content of the message is precisely the thermodynamic entropy (modulo the factor of the Boltzmann constant and the change from base $e$ to base $2$ in the logarithm.

In quantum mechanics, the concept of the classical probability distribution is replaced by the density matrix
\be
\rho = \sum_j P_j |\psi_j\rangle\langle\psi_j|.
\ee
Here we imagine a situation in which state $|\psi_j\rangle$ in the Hilbert space is drawn at random from the ensemble with classical probability $P_j$.  Note that the different $|\psi_j\rangle$ are normalized but need not be orthogonal.  The ensemble average of any observable ${\cal O}$ is given by
\be
\langle {\cal O}\rangle = \Tr\{{\cal O}\rho\}.
\label{eq:tracerho}
\ee

 \boxedtext{
  \begin{problem}
  Prove the identity in Eq.~(\ref{eq:tracerho}) by computing the trace in some basis and rearranging the sum on all states so that it becomes an insertion of the identity.  Do not assume that the different $|\psi_j\rangle$ are orthogonal.
\end{problem}
}

The density matrix for a system with a Hilbert space of dimension $M$ is always an $M\times M$ positive semi-definite (i.e. having only zero or positive real eigenvalues) Hermitian matrix with unit trace.  In thermal equilibrium, the density matrix is given by
\be
\rho=\frac{1}{Z}e^{-\beta H}
\ee
with $Z\equiv \Tr e^{-\beta H}$.

For a single spin-1/2, the density matrix has the following convenient general representation
\be
\rho = \frac{1}{2}\sum_{j=0}^3 \langle Q_j\rangle\, Q_j,
\label{eq:singlespinrhorep1}
\ee
where $Q_0=\hat I,Q_1=\sigma^x,Q_2=\sigma^y,Q_3=\sigma^z$ and $\hat I$ is the identity operator.  This can be rewritten
\be
\rho = \frac{1}{2}\left[\hat I + \langle \vec\sigma\rangle\cdot\vec\sigma\right].
\label{eq:singlespinrhorep2}
\ee
Thus the density matrix for a single spin is determined solely by the spin polarization of the ensemble.  The ensemble could represent a large collection of spins (as in NMR) or the average results of measuring a single qubit many times (allowing the qubit to come to equilibrium with its environment before each measurement is repeated).

\boxedtext{
  \begin{problem}
  Prove the identity in Eq.~(\ref{eq:singlespinrhorep2}) by first showing that $\Tr Q_j=2\delta_{j,0}$ and $\Tr Q_jQ_k=2\delta_{j,k}$.
\end{problem}
}

The analog of the Shannon entropy for a classical probability distribution is the von Neuman entropy of the density matrix
\be
S=-\Tr \{\rho \ln_2\rho\}.
\ee
Evaluating the trace in the basis of eigenstates of $\vec m\cdot\vec\sigma$, where $\vec m=\langle\vec\sigma\rangle$ is the Bloch vector (spin polarization), it is straigtforward to show that
\be
S=-\left\{\frac{1+m}{2}\ln_2\left[\frac{1+m}{2}\right]+\frac{1-m}{2}\ln_2\left[\frac{1-m}{2}\right]\right\}
\label{eq:entropyversusm}
\ee
where $m=|\vec m|$.  From this it follows that $S(m)$ decreases monotonically from $S(0)=1$ to $S(1)=0$.  Zero von Neuman entropy means that a system is in a single pure state
\be
\rho=|\phi\rangle\langle\phi|,
\ee
from which it follows that
\be
\rho^2=\rho
\ee
for a pure state.  Equivalently, the density matrix of a pure state has one eigenvalue of unity and $M-1$ zero eigenvalues.

Before turning finally to the concept of entanglement entropy, let us generalize the above results to the case of two qubits.  The density matrix is $4\times 4$ and the analog of Eq.~(\ref{eq:singlespinrhorep1}) is
\be
\rho = \frac{1}{4}\sum_{j,k=0}^3 \langle Q_j^{(1)}Q_k^{(2)}\rangle\, Q_j^{(1)}Q_k^{(2)},
\label{eq:twospinrhorep2}
\ee

\boxedtext{
  \begin{problem}
  Prove the identity in Eq.~(\ref{eq:twospinrhorep2}) by first showing that $\Tr Q_j^{(a)}Q_k^{(b)}=4\delta_{j,k}\delta_{a,b}$.
\end{problem}
}

\begin{figure}[ht]
\centerline{\psfig{file=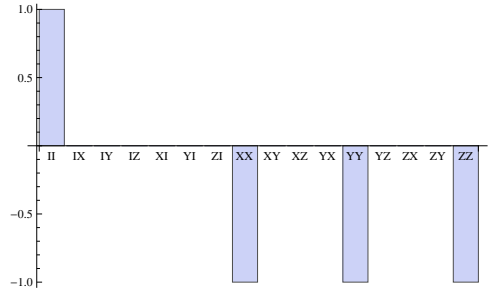,width=2.5in}}
\caption{`Pauli bar plot' of one and two spin correlators in the Bell state $|B_0\rangle$.}
\label{fig:PaulibarplotBellState0}
\end{figure}

We can get a clear picture of the Bell states by examining the two-spin correlators $\langle Q_j^{(1)}Q_k^{(2)}\rangle$ in the so-called `Pauli bar plot' for the state $|B_0\rangle$ shown in  Fig.~(\ref{fig:PaulibarplotBellState0}). We see that all the single-spin expectation values vanish.  Because of the entanglement, each spin is on average totally unpolarized.  Yet three of the two-spin correlators, $XX,YY,ZZ$, are all -1 indicating that the two spins are pointing in exactly opposite directions.  This is becasue $|B_0\rangle$ is the rotationally invariant spin-singlet state.

With this background, we are now ready to tackle the concept of entanglement entropy.  Consider product state $|\uparrow\uparrow\rangle$ with density matrix
\be
\rho=|\uparrow\uparrow\rangle\langle\uparrow\uparrow|.
\ee
Suppose now that we only had access to the first qubit and not the second.  We can compute the expectation value of any observable associated with the first qubit using Eq.~(\ref{eq:tracerho}), but it will be convenient to carry out the trace in two steps, first over the states of the second qubit state and then over those of the first.  We can do this by defining the so-called `reduced density matrix' for the first spin
\be
\bar\rho_1\equiv \Tr_2\rho.
\ee
This is an ordinary $2\times 2$ density matrix for the first spin.  In this simple case
\be
\bar\rho= \Tr_2|\uparrow\uparrow\rangle\langle\uparrow\uparrow| = {|\uparrow\rangle_1} {_1\langle\uparrow|}
\ee
corresponding to a pure state with the first qubit in the up state.   In a product state, individual qubits can be viewed as being separately in their own pure states.

Consider now what happens when we compute the reduced density matrix for the first qubit in the entangled Bell state $|B_0\rangle$.  This is a pure state with zero entropy.  However computation of the reduced density matrix yields
\be
\bar\rho= \frac{1}{2}\left\{|\uparrow\rangle\langle\uparrow|+|\downarrow\rangle\langle\downarrow|\right\}
=\frac{1}{2}\left(\begin{array}{cc}1&0\\0&1\end{array}\right),
\label{eq:barrhoB_0}
\ee
which is an equal statistical mixture of up and down spin.  This has the maximum possible entropy of 1 bit.  Similarly, an observer examining only the second qubit would see an impure state with the maximum possible entropy.  This seems to be at odds with the overall state of the combined system having zero entropy and being in a pure state.  Indeed if this were a classical system it would be impossible. To see why consider two identical statistical mechanical systems with entropy $S_1=S_2$.  Normally we expect the entropy of the universe to simply be the sum $S_1+S_2$.  Suppose however that the microstate of system 2 is perfectly correlated with (e.g. identical to) that of system 1.  Thus if we learn which microstate system 1 is in, we gain no new information when we look at the microstate of system 2 since we can predict it ahead of time.  The entropy of the combined system is just $S_1$, not $2S_1$.  This reasoning leads us to the realization that classically the entropy of a composite system cannot be less than the entropy of the component with the largest entropy.

\boxedtext{
\begin{problem}
Derive Eq.~(\ref{eq:barrhoB_0}).
\end{problem}}

The quantum result derived above, that each component has finite entropy but the overall system has zero entropy is in clear contradiction to the classical result\footnote{A situation that Charles Bennett has summarized by stating that `A classical house is at least as dirty as its dirtiest room, but a quantum house can be dirty in every room, yet still perfectly clean overall.'}.  Evidently there is a kind of `negative entropy' associated with the non-classical correlations which cancels out the entropy of the separate components in an entangled system.  The `entanglement entropy' of a bipartite system  which is in a pure state is defined by von Neumann entropy of the reduced density matrix of each of the subsystems
\be
S_\mathrm{E}=S(\bar\rho_1)=S(\bar\rho_2).
\ee
For the Bell states, $S_\mathrm{E}=1\,$bit, while for any product state, $S_\mathrm{E}$ vanishes.   Notice that the fact that we found the polarization of the individual qubits in the Bell state vanishes is consistent with Eq.~(\ref{eq:entropyversusm}) which says that the entropy is a maximum when the polarization vanishes.

Entanglement entropy is one quantitave measure of the entanglement between two halves of a bipartite system and in recent years has (along with the `entanglement spectrum'\cite{HaldaneEntanglementSpectrum}) become a valuable theoretical tool in the analysis of correlations in many-body systems\cite{EisertRMP,HorodeckiRMP,many-body-quantum-info,Verstraete} and has even forged connections between condensed matter physics and quantum gravity\cite{SachdevReview}.  It is also clear from the discussion above that if a qubit in a quantum computer becomes entangled with an external reservoir due to spurious coupling, the coherence needed to carry out computations is lost.

Entanglement also enters the modern picture of quantum measurement\cite{ToddBrunQuantumTrajectories,KurtJacobsContinuousQuantumMeasurement,Clerk2008,HarocheRaimondcQEDBook}.  In the Stern-Gerlach experiment, one does not measure the spin directly.  Rather the magnetic field gradient entangles spin with position, and then one measures the position.  From the position one infers the spin projection.  Suppose that we start with an initial wave function in which the spin is pointing in some arbitrary direction
\be
|\Psi\rangle = [\alpha|\uparrow\rangle+\beta|\downarrow\rangle]\Phi(\vec r),
\ee
where $\Phi$ is the initial spatial wave packet of the spin as it approaches the measurement apparatus.  After passing through the Stern-Gerlach magnet (which has its field gradient in the $z$ direction), the up spin component's trajectory is deflected upward and the down spin component is deflected downward resulting in
\be
|\Psi\rangle = [\alpha|\uparrow\rangle\Phi_\uparrow(\vec r)+\beta|\downarrow\rangle\Phi_\downarrow(\vec r)].
\ee
If the deflection is sufficiently strong that the two spatial states cease to have common support ($\Phi_\uparrow(\vec r)\Phi_\downarrow(\vec r)=0$ for all $\vec r$), then a measurement of position unambiguously determines the spin projection.\footnote{Technically to achieve perfect distinguishability, one only needs a weaker condition, namely that the two spatial states are orthogonal.  However in this more general case you might need to measure some more complex operator to fully distinguish the two measurement `pointer states.'}

Notice that prior to the measurement, the reduced density matrix for the spin degree of freedom is simply
\be
\bar\rho = [\alpha|\uparrow\rangle+\beta|\downarrow\rangle][\alpha^*\langle\uparrow|+\beta^*\langle\downarrow|]=
\left(\begin{array}{c}\alpha\\ \beta\end{array}\right)\left(\begin{array}{cc}\alpha^*&\beta^*\end{array}\right)
=\left(\begin{array}{cc}|\alpha|^2&\alpha\beta^*\\\alpha^*\beta&|\beta|^2\end{array}\right).
\ee
The non-zero off-diagonal terms (`cohences') imply that there will be transverse components of the spin polarization in addition to the $z$ component
\begin{eqnarray}
\langle \sigma^x\rangle &=& \Tr\sigma^x\bar\rho=2\,\mathrm{Re}[\alpha^*\beta]\\
\langle \sigma^y\rangle &=& \Tr\sigma^y\bar\rho=2\,\mathrm{Im}[\alpha^*\beta]\\
\langle \sigma^z\rangle &=& \Tr\sigma^z\bar\rho=|\alpha|^2-|\beta|^2.
\end{eqnarray}
After the spin passes through the magnet, the full density matrix for spin and position is
\be
\rho(\vec r, \vec r^{\,\prime}) = \langle\vec r|\rho|\vec r^{\,\prime}\rangle=\left(\begin{array}{lr}|\alpha|^2|\Phi_\uparrow(\vec r)\Phi^*_\uparrow(\vec r^{\,\prime})&\,\alpha\beta^*\Phi_\uparrow(\vec r)\Phi^*_\downarrow(\vec r^{\,\prime})\\ \alpha^*\beta\Phi^*_\uparrow(\vec r)\Phi_\downarrow(\vec r^{\,\prime})&\,|\beta|^2\Phi_\downarrow(\vec r)\Phi^*_\downarrow(\vec r^{\,\prime})\end{array}\right).
\ee
and the unconditional reduced density matrix for the spin (tracing out the position by integrating over $\vec r=\vec r^{\,\prime}$ and therefore ignoring the measurement result!) yields
\be
\bar\rho_\mathrm{uc}=\left(\begin{array}{cc}|\alpha|^2&\alpha\beta^*\langle\Phi_\uparrow|\Phi_\downarrow\rangle\\\alpha^*\beta\langle\Phi_\downarrow|\Phi_\uparrow\rangle&|\beta|^2\end{array}\right).
\label{eq:barrhouc1}
\ee
We see that as the measurement pointer states become orthogonal, the off-diagonal elements decay away and the density matrix becomes impure.  This phenomenon is referred to as measurement-induced dephasing.  One can show that the rate of dephasing is directly equivalent to the rate of information gain in the measurement.\cite{Clerk2008} (We emphasize again however that we are here referring to the ensemble average over the measurement results which ignores the particular measurement result.)

Instead of averaging over all measurement results, we can ask the following question:  What is the spin density matrix conditioned on the measurement result being $\vec R$?  This is given by
\be
\bar\rho_\mathrm{c}(\vec R) = \frac{1}{P(\vec R)}\langle\vec R|\rho|\vec R\rangle=\frac{1}{P(\vec R)}\left(\begin{array}{lr}|\alpha|^2|\Phi_\uparrow(\vec R)\Phi^*_\uparrow(\vec R)&\,\alpha\beta^*\Phi_\uparrow(\vec R)\Phi^*_\downarrow(\vec R)\\ \alpha^*\beta\Phi^*_\uparrow(\vec R)\Phi_\downarrow(\vec R)&\,|\beta|^2\Phi_\downarrow(\vec R)\Phi^*_\downarrow(\vec R)\end{array}\right),
\ee
where we have introduced the factor
\be
P(\vec R) \equiv |\alpha|^2 |\Phi_\uparrow(\vec R)|^2 + |\beta|^2|\Phi_\downarrow(\vec R)|^2,
\ee
to satisfy the probability normalization condition $\Tr\bar\rho_\mathrm{c}(\vec R)=1$.  Notice that $P(\vec R)$ is precisely the probability density that the position measurement (unconditioned on the spin state) will yield the value $\vec R$.  This normalization factor also makes sense because if we ensemble average over all measurement results (weighted by the probability of occurrence of each measurement result) we correctly recover the unconditional reduced density matrix of Eq.~(\ref{eq:barrhouc1})
\be
\bar\rho_\mathrm{uc}=\int d^3R\, P(\vec R) \bar\rho_\mathrm{c}(\vec R),
\label{eq:barrhouc2}
\ee
because the normalization factor cancels out.

Remarkably, the reduced density matrix conditioned on the measurement result corresponds to a pure state.  This can be easily seen by noting that we can write the conditional density matrix in terms of a conditional state vector (wave function)
\be
\bar\rho_\mathrm{c} = |\Psi_\mathrm{c}\rangle\langle\Psi_\mathrm{c}|,
\ee
where
\be
|\Psi_\mathrm{c}\rangle \equiv \frac{1}{\sqrt{P(\vec R)}}\left[\alpha_\mathrm{c}|\uparrow\rangle + \beta_\mathrm{c}|\downarrow\rangle     \right],
\ee
and the new coefficients depend on the particular measurement result, $\vec R$
\begin{eqnarray}
\alpha_\mathrm{c} &=& \alpha\Phi_\uparrow(\vec R)\\
\beta_\mathrm{c} &=& \beta\Phi_\downarrow(\vec R).
\end{eqnarray}
The change in the state induced by the act of measurement is known as the `back action.'  This change in state modifies the spin polarization
\begin{eqnarray}
\langle \sigma^x\rangle &=&\frac{2}{P(\vec R)}\,\mathrm{Re}[\alpha_\mathrm{c}^*\beta_\mathrm{c}]\\
\langle \sigma^y\rangle &=&\frac{2}{P(\vec R)}\,\mathrm{Im}[\alpha_\mathrm{c}^*\beta_\mathrm{c}]\\
\langle \sigma^z\rangle &=&\frac{1}{P(\vec R)}\left[\alpha_\mathrm{c}|^2-|\beta_\mathrm{c}|^2\right].
\end{eqnarray}
For a weak measurement $\Phi_\uparrow$ and $\Phi_\downarrow$ do not differ by much and the backaction is small. Experimental progress in developing nearly ideal quantum limited amplifiers is such that it is now possible to directly observe this backaction and how it varies with the strength of the measurement.\cite{HatridgeBackaction}  For a strong measurement, $\Phi_\uparrow$ and $\Phi_\downarrow$ are fully separated and have no common support.  Thus $\Phi_\uparrow(\vec R)\Phi_\downarrow(\vec R)=0$ and we have complete collapse of the spin polarization to
\begin{eqnarray}
\langle \sigma^x\rangle &=&0\\
\langle \sigma^y\rangle &=&0\\
\langle \sigma^z\rangle &=&\pm 1.
\end{eqnarray}
where the sign of $\langle\sigma^z\rangle$ is determined by $\vec R$ and is positive with probability $|\alpha|^2$ and negative with probability $|\beta|^2$.

Our discussion so far has been of the entanglement between a spin and the measurement apparatus pointer.  We have seen that a strong measurement of the pointer variable (which in the example above collapses the pointer variable to a particular value of $\vec R$) leads to a weak or strong back action on the spin which partially or completely collapses it to a definite state depending on the degree of entanglement (the measurement strength).  We turn now to further consideration of the `spooky' correlations in entangled states.

We have already seen for the Bell state $|B_0\rangle$ that the components of the two spins are perfectly anti-correlated.  Suppose now that Alice prepares two qubits in this Bell state and then sends one of the two qubits to Bob who is far away (say one light-year).  Alice now chooses to make make a measurement of her qubit projection along some arbitrary axis $\hat n$.  For simplicity let us say that she chooses the $\hat z$ axis.  Let us further say that her measurement result is $-1$.  Then she immediately knows that if Bob chooses to measure his qubit along the same axis, his measurement result will be the opposite, $+1$.  It seems that Alice's measurement has collapsed the state of the spins from $|B_0\rangle$ to $|\downarrow\uparrow\rangle$.  This `spooky action at a distance' in which Alice's measurement seems to instantaneously change Bob's distant qubit was deeply troubling to Einstein.\cite{EPR}

Upon reflection one can see that this effect cannot be used (in violation of special relativity) for superluminal communication.  Even if Alice and Bob and agreed in advance on what axis to use for measurements, Alice has no control over her measurement result and so cannot use it to signal Bob.  It is true that Bob can immediately find out what Alice's measurement result was, but this does not give Alice the ability to send a message.   In fact, suppose that Bob's clock was off and he accidentally made his measurement slightly before Alice. Would either he or Alice be able to tell?  The answer is no, because each would see a random result just as expected.  This must be so because in special relativity, the concept of simultaneity is frame-dependent and not universal.

\begin{figure}[ht]
\centerline{\psfig{file=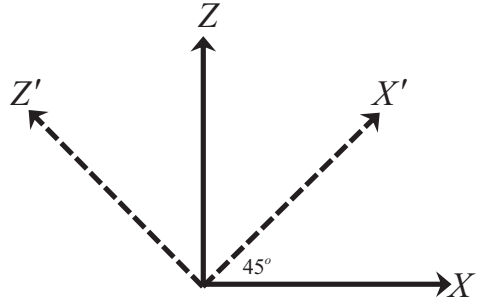,width=2.5in}}
\caption{Measurement axes used by Alice (solid lines) and Bob (dashed lines) in establishing the Clauser-Horn-Shimoni-Holt (CHSH) inequality.}
\label{fig:CHSHaxes}
\end{figure}

Things get more interesting when Alice and Bob choose different measurement axes.  Einstein felt that quantum mechanics must somehow not be a complete description of reality and that there might be `hidden variables' which if they could be measured would remove the probabilistic nature of the quantum description.  However in 1964 John S. Bell proved a remarkable inequality\cite{BellNoGo} showing that when Alice and Bob use certain different measurement axes, the correlations between their measurement results are stronger than any possible classical correlation that could be induced by (local) hidden variables.  Precision experimental measurements which violate the Bell inequalities are now the strongest proof that quantum mechanics is correct and that local hidden variable theories are excluded.   Perhaps the simplest way to understand this result is to consider the CHSH inequality developed by Clauser, Horn, Shimoni and Holt\cite{CHSH} following Bell's ideas.  Consider the measurement axes shown in Fig.~(\ref{fig:CHSHaxes}).  The experiment consists of many trials of the following protocol. Alice and Bob share a pair of qubits in an entangled state.   Alice randomly chooses to measure the first qubit using $X$ or $Z$ while Bob randomly chooses to measure the second qubit using $X^\prime$ or $Z^\prime$.  After many trials (each starting with a fresh copy of the entangled state), Alice and Bob compare notes on their measurement results and compute the following correlation function:
\be
S=\langle XX^\prime\rangle + \langle ZZ^\prime\rangle - \langle XZ^\prime\rangle + \langle ZX^\prime\rangle,
\ee
which can be rewritten
\be
S=\langle (X+Z)X^\prime\rangle - \langle (X-Z)Z^\prime\rangle.
\ee
Alice and Bob note that their measurement results are random variables which are always equal to either $+1$ or $-1$. In a particular trial Alice chooses randomly to measure either $X$ or $Z$.  Surely however, the variable not measured still has a value of either $+1$ or $-1$.  If this is true, then either $X=Z$ or $X=-Z$.  Thus either $X+Z$ vanishes or $X-Z$ vanishes in any given realization of the random variables. The combination that does not vanish is either $+2$ or $-2$.  Hence it follows immediately that $S$ is bounded by the CHSH inequality
\be
-2\le S\le +2.
\ee

It turns out however that the quantum correlations between the two spins in the entangled pair violate this classical bound.  They are stronger than can ever be possible in any classical local hidden variable theory.  Because $\vec \sigma$ is a vector we can resolve its form in one basis in terms of the other via
\begin{eqnarray}
\sigma^{x'}&=&\frac{1}{\sqrt{2}}[\sigma^z+\sigma^x]\\
\sigma^{z'}&=&\frac{1}{\sqrt{2}}[\sigma^z-\sigma^x].
\end{eqnarray}
Thus we can express $S$ in terms of the 'Pauli bar' correlations
\be
S=\frac{1}{\sqrt{2}}\left[\langle XX+XZ+ZX+ZZ\rangle -\langle XZ-XX-ZZ+ZX\rangle  \right].
\ee
For Bell state $|B_0\rangle$, these correlations are shown in Fig.~(\ref{fig:PaulibarplotBellState0}) and yield
\be
S=-2\sqrt{2},
\ee
in clear violation of the CHSH inequality.  Strong violations of the CHSH inequality are routine in modern experiments.\footnote{Although strictly speaking, there are loopholes associated with imperfections in the detectors and the fact that Alice and Bob typically do not have a space-like separation.} This teaches us that in our quantum world, observables do not have values if you do not measure them.  There are no hidden variables which determine the random values.  The unobserved spin components simply do not have values.  Recall that $\sigma^x$ and $\sigma^z$ are incompatible observables and when we choose to measure one, the other remains not merely unknown, but unknowable.

It is ironic that Einstein's arguments that quantum mechanics must be incomplete because of the spooky properties of entanglement have led to the strongest experimental tests verifying the quantum theory and falsifying all local hidden variable theories.  We are forced to give up the idea that physical observables have values before they are observed.

\subsection{Quantum Dense Coding}
We saw above that quantum correlations are strong enough to violate certain classical bounds, but the spooky action at a distance seemed unable to help us send signals.  Actually it turns out that by using a special `quantum dense coding' protocol, we can use entanglement to help us transmit information in a way that is impossible classically.\cite{quantumdensecoding}  As background let us recall that for a single qubit there are only two orthogonal states which are connected by a rotation of the spin through an angle of $\pi$.    For rotation by $\pi$ around the $y$ axis the unitary rotation operator is
\be
R^y_\pi = e^{-i\frac{\pi}{2}\sigma^y}= -i\sigma^y,
\ee
and we map between the two orthogonal states
\begin{eqnarray}
R^y_\pi|\uparrow\rangle &=& +|\downarrow\rangle\\
R^y_\pi|\downarrow\rangle &=& -|\uparrow\rangle
\end{eqnarray}
while any other rotation angle simply produces linear general combinations of the two basis states.  For example, for rotation by $\pi/2$ we have
\be
R^y_\frac{\pi}{2} = e^{-i\frac{\pi}{4}\sigma^y}= \frac{1}{\sqrt{2}}[1-i\sigma^y],
\ee
and we have
\begin{eqnarray}
R^y_\frac{\pi}{2}|\uparrow\rangle &=& \frac{1}{\sqrt{2}}[|\uparrow\rangle + |\downarrow\rangle]=|\rightarrow\rangle\\
R^y_\frac{\pi}{2}|\downarrow\rangle &=& -\frac{1}{\sqrt{2}}[|\uparrow\rangle - |\downarrow\rangle]=-|\leftarrow\rangle.
\end{eqnarray}
These are linearly independent of the starting state but never orthogonal to it except for the special case of rotation by $\pi$.

The situation is very different for two-qubit entangled states.  We take as our basis the four orthogonal Bell states. Suppose that Alice prepares the Bell state $|B_0\rangle$ and sends one of the qubits to Bob who is in a distant location.  Using a remarkable protocol called quantum dense coding\cite{quantumdensecoding}, Alice can now send Bob two classical bits of information by sending him the remaining qubit.  The protocol relies on the amazing fact that Alice can transform the initial Bell state into any of the others by purely local operations on her remaining qubit without communicating with Bob.  The four possible unitary operations Alice should perform are $I,X,iY,Z$ which yield
\begin{eqnarray}
I|B_0\rangle  &=& |B_0\rangle\\
Z|B_0\rangle  &=& |B_1\rangle\\
X|B_0\rangle  &=& |B_2\rangle\\
iY|B_0\rangle &=& |B_3\rangle.
\end{eqnarray}
It seems somehow miraculous that without touching Bob's qubit, Alice can reach all four orthogonal states by merely rotating her own qubit.  This means that there are four possible two-bit messages Alice can send by associating each with one of the four operations $I,X,iY,Z$ according to the following
%\begin{table}
\begin{center}
\begin{tabular}{|c|c|}
\hline
message&operation\\
\hline
00&X\\
%\cline
01&I\\
%\cline
10&iY\\
%\cline
11&Z\\
\hline
\end{tabular}
%\end{table}
\end{center}
After encoding her message by carrying out the appropriate operation on her qubit, Alice physically transmits her qubit to Bob.  Bob then makes a joint measurement (details provided further below) on the two qubits which tells him which of the four Bell states he has and thus recovers two classical bits of information even though Alice sent him only one quantum bit after deciding what the message was.  The pre-positioning of the entangled pair has given them a resource which doubles the number of classical bits that can be transmitted with one (subsequent) quantum bit.  Of course in total Alice transmitted two qubits to Bob.  The key point is that the first one was sent in advance of deciding what the message was.  How weird is that!?

This remarkable protocol sheds considerable light on the concerns that Einstein raised in the EPR paradox.\cite{EPR}  It shows that the special correlations in Bell states can be used to communicate information in a novel and efficient way by `prepositioning' entangled pairs shared by Alice and Bob.  However causality and the laws of special relativity are not violated because Alice still has to physically transmit her qubit(s) to Bob in order to send the information.

\begin{figure}[ht]
\centerline{\psfig{file=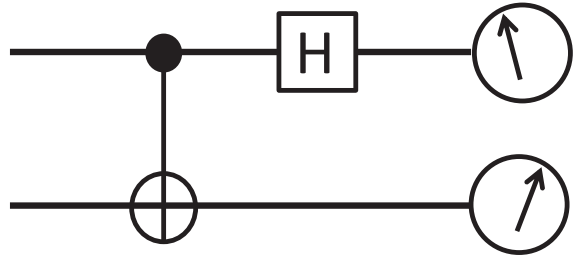,width=3.0in}}
\caption{Bell measurement circuit with a CNOT and Hadamard gate.  This circuit permits measurement of which Bell state a pair of qubits is in by mapping the states to the basis of eigenstates of $\sigma^z_1$ and $\sigma^z_2$.}
\label{fig:BellMeasurementCircuit}
\end{figure}

The above protocol requires Bob to determine which of the four Bell states he has.  The quantum circuit shown in Fig.~(\ref{fig:BellMeasurementCircuit}) uniquely maps each of the Bell states onto one of the four computational basis states (eigenstates of $Z_1$ and $Z_2$).  The first symbol indicates the CNOT gate which flips the target qubit (in this case qubit 2) if and only if the control qubit (qubit 1) is in the excited state. This can be explicitly written in matrix form as
\be
\mathrm{CNOT}_{12} = \frac{\sigma^0_1+\sigma^z_1}{2}\sigma^x_2+\frac{\sigma^0_1-\sigma^z_1}{2}\sigma^0_2,
\ee
where $\sigma^0_j$ is the identity operator for qubit $j$.  The second gate in the circuit (denoted H) is the Hadamard gate acting on the first qubit
\be
\mathrm{H}=\frac{1}{\sqrt{2}}[\sigma^z+\sigma^x].
\ee
The Hadamard gate obeys $\mathrm{H}^2=1$ and interchanges the $X$ and $Z$ components of a spin:
\begin{eqnarray}
H|\uparrow\rangle &=& |\rightarrow\rangle\\
H|\downarrow\rangle &=& |\leftarrow\rangle\\
H|\rightarrow\rangle &=& |\uparrow\rangle\\
H|\leftarrow\rangle &=& |\downarrow\rangle
\end{eqnarray}

\boxedtext{\begin{problem}
Prove the following identities for the circuit shown in Fig.~(\ref{fig:BellMeasurementCircuit})
\begin{eqnarray}
\mathrm{H}_1\, \mathrm{CNOT}_{12} |B_0\rangle &=& +|ge\rangle\\
\mathrm{H}_1\, \mathrm{CNOT}_{12} |B_1\rangle &=& +|ee\rangle\\
\mathrm{H}_1\, \mathrm{CNOT}_{12} |B_2\rangle &=& -|gg\rangle\\
\mathrm{H}_1\, \mathrm{CNOT}_{12} |B_3\rangle &=& -|eg\rangle
\end{eqnarray}
\end{problem}}

Once Bob has mapped the Bell states onto unique computational basis states, he makes a separate measure of the state of each qubit thereby gaining two bits of classical information and effectively reading the message Alice has sent. Note that the overall sign in front of the basis states produced by the circuit is irrelevant and not observable in the measurement process.  Also note that to create Bell states in the first place, Alice can simply run the circuit in Fig.~(\ref{fig:BellMeasurementCircuit}) backwards.

\subsection{No-Cloning Theorem Revisited}

Now that we have understood the EPR paradox and communication via quantum dense coding, we can gain some new insight into the no-cloning theorem.  It turns out that if cloning of an unknown quantum state were possible, then we could use an entangled pair to communicate information at superluminal speeds in violation of special relativity.  Consider the following protocol.  Alice and Bob share an entangled Bell pair in state $|B_0\rangle$.  Alice chooses to measure her qubit in either the $Z$ basis or the $X$ basis.  The choice she makes defines one classical bit of information.   The result of the measurement collapses the entangled pair into a simple product state. If Alice chooses the $Z$ basis for her measurement, then Bob's qubit will be either $|\uparrow\rangle$ or $|\downarrow\rangle$.  If Alice chooses to measure in the $X$ basis, then Bob's qubit will be either $|\rightarrow\rangle$ or $|\leftarrow\rangle$.  Bob can distinguish these cases by cloning his qubit to make many copies.  If he measures a large number of copies in the $Z$ basis and always gets the same answer, he knows that his qubit is almost certainly in a $Z$ eigenstate.  If even a single measurement result is different from the first, he knows his qubit cannot be in a $Z$ eigenstate and so must be in an $X$ eigenstate.  (Of course he could also measure a bunch of copies in the $X$ basis and gain the same information.)  This superluminal communication would violate special relativity and hence cloning must be impossible.

In fact, cloning would make it possible for Alice to transmit an unlimited number of classical bits using only a single Bell pair.  Alice could choose an arbitrary measurement axis $\hat n$.  The specification of $\hat n$ requires two real numbers (the polar and azimuthal angles).  It would take a very large number of bits to represent these real numbers to some high accuracy.  Now if Bob can make an enormous number of copies of his qubit, he can divide the copies in three groups and measure the vector spin polarization $\langle \vec\sigma\rangle$ to arbitrary accuracy.  From this he knows the polarizaton axis $\hat n=\pm \langle \vec \sigma\rangle$ Alice chose (up to an unknown sign since he does not know the sign of Alice's measurement result for $\hat n\cdot\sigma$).  Hence Bob has learned a large number of classical bits of information.   The accuracy (and hence the number of bits) is limited only by the statistical uncertainties resulting from the fact that his individual measurement results can be random, but these can be reduced to an arbitrarily low level with a sufficiently large number of copies of the state.

\subsection{Quantum Teleportation}

We noted earlier than even though it is impossible to clone an unknown quantum state, it is possible for Alice to teleport it to Bob as long as her copy of the original is destroyed in the process\cite{teleportation}.  Just as for quantum dense coding, teleportation protocols also take advantage of the power of `pre-positioned' entangled pairs.  However unlike quantum dense coding where Alice ultimately sends her qubit to Bob, teleportation only requires Alice to send two classical bits to Bob.  A simple protocol is as follows:  Alice creates a $|B_0\rangle$ Bell state and sends one of the qubits to Bob.  Alice has in her possession an additional qubit in an unknown state
\be
|\psi\rangle = \alpha |g\rangle + \beta |e\rangle
\ee
which she wishes to transmit to Bob.  Alice applies the Bell state determination protocol illustrated in Fig.~(\ref{fig:BellMeasurementCircuit}) to determine the joint state of the unknown qubit and her half of the Bell pair she shares with Bob. She then transmits two classical bits giving her measurement result to Bob.  To see how Bob is able to reconstruct the initial state, note that we can rewrite the initial state of the three qubits in the basis of Bell states for the two qubits that Alice will be measuring as follows
\begin{eqnarray}
|\Psi\rangle &=& [\alpha|e\rangle + \beta|g\rangle]|B_0\rangle\\
&=& \frac{1}{2}|B_0\rangle[ - \alpha|e\rangle - \beta|g\rangle]\nonumber\\
&+& \frac{1}{2}|B_1\rangle[ - \alpha|e\rangle + \beta|g\rangle]\nonumber\\
&+& \frac{1}{2}|B_2\rangle[ - \beta|e\rangle - \alpha|g\rangle]\nonumber\\
&+& \frac{1}{2}|B_3\rangle[ + \beta|e\rangle - \alpha|g\rangle]
\end{eqnarray}
From this representation we see that when Alice tells Bob which Bell state she found, Bob can find a local unitary operation to perform on his qubit to recover the original unknown state (up to an irrelevant overall sign).  The appropriate operations are
\begin{center}
\begin{tabular}{|c|c|}
\hline
Alice's Bell state& Bob's operation\\
\hline
$|B_0\rangle$&X\\
%\cline
$|B_1\rangle$&-iY\\
%\cline
$|B_2\rangle$&I\\
%\cline
$|B_3\rangle$&Z\\
\hline
\end{tabular}
\end{center}

\section{Quantum Error Correction}

Now that we understand entanglement, we are in a position to tackle quantum error correction.

To overcome the deleterious effects of electrical noise, cosmic rays and other hazards, modern digital computers rely heavily on error correcting codes to store and correctly retrieve vast quantities of data.  Classical error correction works by introducing extra bits which provide redundant encoding of the information.  Error correction proceeds by measuring the bits and comparing them to the redundant information in the auxiliary bits.  Another benefit of the representation of information as discrete bits (with 0 and 1 corresponding to a voltage for example) is that one can ignore small noise voltages.  That is, $V=0.99$ volts can be safely assumed to represent 1 and not 0.

All classical (and quantum) error correction codes are based on the assumption that the hardware is good enough that errors are rare.  The goal is to make them even rarer. For classical bits there is only one kind of error, namely the bit flip which maps 0 to 1 or vice versa.  We will assume for simplicity that there is probability $p\ll 1$ that a bit flip error occurs, and that error occurrences are uncorrelated among different bits.
One of the simplest classical error correction codes to understand involves repetition and majority rule.  Suppose we have a classical bit carrying the information we wish to protect from error and we have available two ancilla bits (also subject to errors).  The procedure consists copying the state of the first bit into the two ancilla bits.  Thus a `logical' 1 is represented by three `physical' bits in state 111, and a `logical' 0 is represented by three `physical' bits in state 000. Any other physical state is an error state outside of the logical state space.  Suppose now that one of the three physical bits suffers an error.  By examining the state of each bit it is a simple matter to identify the bit which has flipped and is not in agreement with the `majority.'  We then simply flip the minority bit so that it again agrees with the majority.  This procedure succeeds if the number of errors is zero or one, but it fails if there is more than one error.  Of course since we have replaced one imperfect bit with three imperfect bits, this means that the probability of an error occurring has increased considerably. For three bits the probability $P_n$ of $n$ errors is given by
\begin{eqnarray}
P_0&=&(1-p)^3\\
P_1&=&3p(1-p)^2\\
P_2&=&3p^2(1-p)\\
P_3&=&p^3.
\end{eqnarray}
Because our error correction code only fails for two or more physical bit errors the error probability for our logical qubit is
\be
p_\mathrm{logical}=P_2+P_3 = 3p^2 - 2p^3,
\ee
If $p <1/2$, then the error correction scheme reduces the error rate (instead of making it worse).  If for example $p=10^{-6}$, then $p_\mathrm{logical}\sim3\times 10^{-12}$.  Thus the lower the raw error rate, the greater the improvement.  Note however that even at this low error rate, a petabyte ($8\times 10^{15}$ bit) storage system would have on average 24,000 errors.  Futhermore, one would have to buy three petabytes of storage since 2/3 of the disk would be taken up with ancilla bits!

We are now ready to enter the remarkable and magic world of quantum error correction.  Without quantum error correction, quantum computation would be impossible and there is a sense in which the fact that error correction is possible is even more amazing and counter intuitive than the fact of quantum computation itself.  Naively, it would seem that quantum error correction is completely impossible.  The no-cloning theorem does not allow us to copy an unknown state of a qubit onto ancilla qubits.  Furthermore, in order to determine if an error has occurred, we would have to make a measurement, and the back action (state collapse) from that measurement would itself produce random unrecoverable errors.

%\section{Quantum Error Correction}

Part of the power of a quantum computer derives from its analog character--quantum states are described by continuous real (or complex) variables.  This raises the specter that noise and small errors will destroy the extra power of the computer just as it does for classical analog computers.  Remarkably, this is {\em not} the case!  This is because the quantum computer also has characteristics that are digital.  Recall that any measurement of the state of qubit always yields a binary result.  Amazingly this makes it possible to perform quantum error correction and keep the calculation running even on an imperfect and noisy computer.  In many ways, this discovery (by Shor \cite{Shor1995,Shor1996} and by Steane \cite{Steane1996a,Steane1996b}) is even more profound and unexpected than the discovery of efficient quantum algorithms that work on ideal computers.  For an introduction to the key concepts behind quantum error correction and fault-tolerant quantum computation, see the reviews by Raussendorf\cite{Raussendorf} and by Gottesman\cite{GottesmanThesis,GottesmanIntro}.

It would seem obvious that quantum error correction is impossible because the act of measurement to check if there is an error would collapse the state, destroying any possible quantum superposition information.  Remarkably however, one can encode the information in such a way that the presence of an error can be detected by measurement, and if the code is sufficiently sophisticated, the error can be corrected, just as in classical computation.  Classically, the only error that exists is the bit flip.  Quantum mechanically there are other types of errors (e.g. phase flip, energy decay, erasure channels, etc.).  However codes have been developed\cite{Shor1995,Shor1996,Steane1996a,Steane1996b,GottesmanThesis,GottesmanIntro,MikeandIke} (using a minimum of 5 qubits) which will correct all possible quantum errors.  By concatenating these codes to higher levels of redundancy, even small imperfections in the error correction process itself can be corrected.  Thus quantum superpositions can in principle be made to last essentially forever even in an imperfect noisy system.  It is this remarkable insight that makes quantum computation possible.  Many other ideas have been developed to reduce error rates.  Alexei Kitaev at Caltech in particular has developed novel theoretical ideas for topologically protected qubits which are impervious to local perturbations \cite{Kitaev,SurfaceCodes}.  The goal of realizing such topologically protected logical qubits is being actively pursued for trapped ions\cite{iontraptopo2,iontraptopo1} and superconducting qubits\cite{Ioffe,SimonNiggToric}.  Certain strongly correlated condensed matter systems may offer the possibility of realizing non-abelian quasiparticle defects which could also be used for topologically protected qubits. For a review of this vast field see Das Sarma et al.\cite{DasSarma}
 %and even quantum gravity\cite{SachdevReview}.

As an entr\'e to this rich field, we will consider a simplified example of one qubit in some state $\alpha|0\rangle+\beta|1\rangle$ plus two ancillary qubits in state $|0\rangle$ which we would like to use to protect the quantum information in the first qubit.  As already noted, the simplest classical error correction code simply replicates the first bit twice and then uses majority voting to correct for (single) bit flip errors.  This procedure fails in the quantum case because the no-cloning theorem prevents replication of an unknown qubit state.  Thus there does not exist a unitary transformation which takes
\begin{equation}
[\alpha|0\rangle+\beta|1\rangle]\otimes|00\rangle \longrightarrow [\alpha|0\rangle+\beta|1\rangle]^{\otimes 3}.
\end{equation}
As was mentioned earlier, this is clear from the fact that the above transformation is not linear in the amplitudes $\alpha$ and $\beta$ and quantum mechanics is linear. One can however perform the repetition code transformation:
\begin{equation}
[\alpha|0\rangle+\beta|1\rangle]\otimes|00\rangle \longrightarrow [\alpha|000\rangle+\beta|111\rangle],
\end{equation}
since this is in fact a unitary transformation.
Just as in the classical case, these three physical qubits form a single logical qubit.  The two logical basis states are
\begin{eqnarray}
|0\rangle_\mathrm{log} &=& |000\rangle\nonumber\\
|1\rangle_\mathrm{log} &=& |111\rangle.
\end{eqnarray}
The analog of the single-qubit Pauli operators for this logical qubit are readily seen to be
\begin{eqnarray}
X_\mathrm{log} &=& X_1X_2X_3\nonumber\\
Y_\mathrm{log} &=& i X_\mathrm{log}Z_\mathrm{log}\nonumber\\
Z_\mathrm{log} &=& Z_1Z_2Z_3.
\label{eq:logops}
\end{eqnarray}
We see that this encoding complicates things considerably because now to do even a simple single logical qubit rotation we have to perform some rather non-trivial three-qubit joint operations.  It is not always easy to achieve an effective Hamiltonian that can produce such joint operations, but this is an essential price we must pay in order to carry out quantum error correction.

It turns out that this simple code cannot correct all possible quantum errors, but only a single type.  For specificity, let us take the error operating on our system to be a single bit flip, either $X_1,X_2$, or $X_3$.  These three together with the identity operator, $I$, constitute the set of operators that produce the four possible error states of the system we will be able to correctly deal with.  Following the formalism developed by Daniel Gottesman\cite{GottesmanThesis,GottesmanIntro,Raussendorf}, let us define two \emph{stabilizer} operators
\begin{eqnarray}
S_1&=&Z_1Z_2\\
S_2 &=& Z_2Z_3.
\end{eqnarray}
These have the nice property that they commute both with each other and with all three of the logical qubit operators listed in Eq.~(\ref{eq:logops}).  This means that they can both be measured simultaneously and that the act of measurement does \emph{not} destroy the quantum information stored in any superposition of the two logical qubit states.   Furthermore they each commute or anticommute with the four error operators in such a way that we can uniquely identify what error (if any) has occurred.   Each of the four possible error states (including no error) is an eigenstate of both stabilizers with the eigenvalues listed in the table below
\begin{center}
\begin{tabular}{|c|c|c|}
\hline
error&$S_1$&$S_2$\\
\hline
$I$&$+1$&$+1$\\
$X_1$&$-1$&$+1$\\
$X_2$&$-1$&$-1$\\
%\cline
$X_3$&$+1$&$-1$\\
\hline
\end{tabular}
\end{center}
Thus measurement of the two stabilizers yields two bits of classical information (called the `error syndrome') which uniquely identify which of the four possible error states the system is in and allows the experimenter to correct the situation by applying the appropriate error operator, $I,X_1,X_2,X_3$ to the system to cancel the original error.

We now have our first taste of fantastic power of quantum error correction.  We have however glossed over some important details by assuming that either an error has occurred or it hasn't (that is, we have been assuming we are in a definite error state).  At the next level of sophistication we have to recognize that we need to be able to handle the possibility of a quantum superposition of an error and no error.  After all, in a system described by smoothly evolving superposition amplitudes, errors can develop continuously.  Suppose for example that the correct state of the three physical qubits is
\be
|\Psi_0\rangle=\alpha|000\rangle + \beta|111\rangle,
\ee
and that there is some perturbation to the Hamiltonian such that after some time there is a small amplitude $\epsilon$ that error $X_2$ has occurred.  Then the state of the system is
\be
|\Psi\rangle = [\sqrt{1-|\epsilon|^2}I + \epsilon X_2]|\Psi_0\rangle.
\label{eq:continouserror}
\ee
(The reader may find it instructive to verify that the normalization is correct.)

What happens if we apply our error correction scheme to this state?  The measurement of each stabilizer will always yield a binary result, thus illustrating the dual digital/analog nature of quantum information processing.  With probability $P_0=1-|\epsilon|^2$, the measurement result will be $S_1=S_2=+1$.
In this case the state collapses back to the original ideal one and the error is removed! Indeed, the experimenter has no idea whether $\epsilon$ had ever even developed a non-zero value.  All she knows is that if there was an error, it is now gone. This is the essence of the Zeno effect in quantum mechanics that repeated observation can stop dynamical evolution.  Rarely however (with probability $P_1=|\epsilon|^2$) the measurement result will be $S_1=S_2=-1$ heralding the presence of an $X_2$ error.  The correction protocol then proceeds as originally described above.  Thus error correction still works for superpositions of no error and one error.  A simple extension of this argument shows that it works for an arbitrary superposition of all four error states.

There remains however one more level of subtlety we have been ignoring.  The above discussion assumed a classical noise source modulating the Hamiltonian parameters.  However in reality, a typical source of error is that one of the physical qubits becomes entangled with its environment.  We generally have no access to the bath degrees of freedom and so for all intents and purposes, we can trace out the bath and work with the reduced density matrix of the logical qubit.  Clearly this is generically not a pure state.  How can we possibly go from an impure state (containing the entropy of entanglement with the bath) to the desired pure (zero entropy) state?  Ordinary unitary operations on the logical qubit preserve the entropy so clearly will not work.  Fortunately our error correction protocol involves applying one of four possible unitary operations \emph{conditioned on the outcome of the measurement of the stabilizers}.  The wave function collapse associated with the measurement gives us just the non-unitarity we need and the error correction protocol works even in this case.  Effectively we have a Maxwell demon which uses Shannon information entropy (from the measurement results) to remove an equivalent amount of von Neumann entropy from the logical qubit!

To see that the protocol still works, we generalize Eq.~(\ref{eq:continouserror}) to include the bath
\be
|\Psi\rangle = [\sqrt{1-|\epsilon|^2}|\Psi_0,\mathrm{Bath}_0\rangle + \epsilon X_2]|\Psi_0,\mathrm{Bath}_2\rangle.
\label{eq:continouserrorbath}
\ee
For example, the error could be caused by the second qubit having a coupling to a bath operator ${\cal O}_2$ of the form
\be
V_2=g\, X_2{\cal O}_2,
\ee
acting for a short time $\epsilon \hbar/g$ so that
\be
|\mathrm{Bath}_2\rangle \approx {\cal O}_2|\mathrm{Bath}_0\rangle.
\ee
Notice that once the stabilizers have been measured, then either the experimenter obtained the result $S_1=S_2=+1$ and the state of the system plus bath collapses to
\be
|\Psi\rangle = |\Psi_0,\mathrm{Bath}_0\rangle,
\ee
or the experimenter obtained the result $S_1=S_2=-1$ and the state collapses to
\be
|\Psi\rangle = X_2|\Psi_0,\mathrm{Bath}_2\rangle.
\ee
Both results yield a product state in which the logical qubit is unentangled with the bath.  Hence the algorithm can simply proceed as before and will work.

\boxedtext{
  \begin{problem}
Compute the Shannon entropy gained by the measurement and show that it is precisely the entanglement entropy which has been removed from the system by the act of measurement.
  \end{problem}
  }

Finally, there is one more twist in this plot.  We have so far described a measurement-based protocol for removing the entropy associated with errors.  There exists another route to the same goal in which purely unitary multi-qubit operations are used to move the entropy from the logical qubit to some ancillae, and then the ancillae are reset to the ground state to remove the entropy.  The reset operation could consist, for example, of putting the ancillae in contact with a cold bath and allowing the qubits to spontaneously and irreversibly decay into the bath. Because the ancillae are in a mixed state with some probability to be in the excited state and some to be in the ground state, the bath ends up in a mixed state containing (or not containing) photons resulting from the decay.  Thus the entropy ends up in the bath.  It is important for this process to work that the bath be cold so that the qubits always relax to the ground state and are never driven to the excited state.  We could if we wished, measure the state of the bath and determine which error (if any) occurred, but in this protocol, no actions conditioned on the outcome of such a measurement are required.  The first error correction circuit for superconducting qubits\cite{YaleQEC} used a minimalist version of such a non-measurement based protocol.

Quantum error correction was first realized experimentally some time ago in NMR\cite{CoryQEC}, then in trapped ions\cite{WinelandQEC}, quantum optics\cite{Pittman2005,Aoki2009} and more recently in superconducting qubits\cite{YaleQEC}.  Despite considerable progress\cite{Knill2001,Boulant2005,Pittman2005,Aoki2009,BlattQECmulti}, we have not yet achieved the goal of a logical qubit which is effectively immortal, but we are moving closer.

%%%%%%%%%%%%%%%%%%%%%%%%%%%%%%
\section{Introduction to Circuit QED}
In the last decade, there has been truly amazing experimental progress in realizing superconducting electrical circuits which operate at microwave frequencies and which behave quantum mechanically.  These circuits exhibit quantized energy levels and can be placed in quantum superpositions of these levels.  An important component of this progress has been the realization that it is extremely useful to apply the ideas of non-linear quantum optics to such circuits.  Quantum electrodynamics is the study of the effect of quantum fluctuations of the electromagnetic field on electrons and atoms.  Quantum fluctuation effects include spontaneous decay (by photon emission) of excited atomic states, Lamb shifts of energy levels, etc.  The spectrum and strength of these quantum fluctuations can be modified by engineering the available electromagnetic modes by placing the system under study in a cavity.  By analogy with this `cavity QED', we will here study `circuit QED'.  In addition to being a novel test bed for non-linear quantum optics in a completely new regime,  circuit QED represents an important architecture for implementing the theoretical ideas about quantum information processing discussed above.

\subsection{Electromagnetic Oscillators}
\begin{figure}[ht]
\centerline{\psfig{file=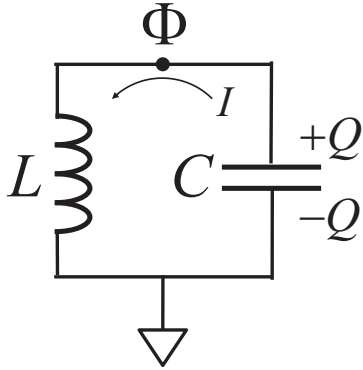,width=2.0in}}
\caption{Simple $LC$ oscillator whose coordinate is the flux $\Phi$ and whose momentum is the charge $Q=C\dot\Phi=CV$.}
\label{fig:LCosc}
\end{figure}

The simplest electrical circuit to understand is the $LC$ oscillator illustrated in Fig.~(\ref{fig:LCosc}).  Let us define the node flux variable
by the time integral of the voltage \cite{MichelLesHouches}
\be
\Phi(t) \equiv \int^t d\tau\, V(\tau).
\ee
The relation $V(t) = \dot\Phi(t)$ shows us (via the Faraday induction relation) that the node flux is indeed the magnetic flux threading the inductor.  The Lagrangian is therefore
\be
L = \frac{1}{2}C\dot\Phi^2 - \frac{1}{2L}\Phi^2,
\ee
In this representation, the kinetic energy term in the Lagrangian represents the electrostatic energy stored in the capacitor and the potential energy term represents the kinetic (magnetic) energy stored in the electron motion through the inductor.

The momentum canonically conjugate to the flux coordinate is given by
\be
Q=\frac{\delta L}{\delta \dot\Phi} = C\dot\Phi=CV,
\ee
and is the charge on the capacitor.  The Hamiltonian is given by
\be
H(Q,\Phi) = Q\dot\Phi - L(\Phi,\dot\Phi)=\frac{Q^2}{2C} + \frac{\Phi^2}{2L}.
\ee
This is the Hamiltonian of a simple harmonic oscillator with `mass' $C$ and `spring constant' $1/L$.  The resonance frequency is therefore $\Omega = \frac{1}{\sqrt{LC}}$ as expected.

We quantize this oscillator by elevating the coordinate and conjugate momentum to operators obeying the canonical commutation relation
\be
[\hat Q,\hat \Phi]=-i\hbar.
\ee
These operators can in turn be represented in terms of raising and lowering operators
\begin{eqnarray}
\hat \Phi &=& \Phi_\mathrm{ZPF} (a + a^\dagger)\label{eq:hatPhi}\\
\hat Q &=& -iQ_\mathrm{ZPF} (a - a^\dagger)
\label{eq:PHIQ}
\end{eqnarray}
where the zero-point fluctuations of the flux and charge are given by
\begin{eqnarray}
\Phi_\mathrm{ZPF} &\equiv& \sqrt{\frac{\hbar Z}{2}}\\
Q_\mathrm{ZPF} &\equiv& \sqrt{\frac{\hbar}{2Z}}
\end{eqnarray}
and $Z\equiv\sqrt{\frac{L}{C}}$ is called the characteristic impedance of the resonator.  Notice that if the characteristic impedance on the order of the quantum of resistance $h/e^2$, then $Q_\mathrm{ZPF}$ is on the order of the electron charge and $\Phi_\mathrm{ZPF}$ is on the order of the flux quantum.

To get a better feeling for the meaning of the characteristic impedance consider the input admittance of the parallel $LC$ resonator
\be
Y(\omega) = j\omega C + \frac{1}{j\omega L}
\ee
where we are using the electrical engineer's convention for the square root of minus one:  j=-i.  This can be rewritten
\be
Y(\omega) = \frac{1}{jZ}\left(\frac{\Omega}{\omega}\right)\left[1-\left(\frac{\omega}{\Omega}\right)^2\right].
\ee
We see that the collective mode frequency is determined by the zero of the admittance as a function of frequency and the slope with which the admittance crosses zero determines the characteristic impedance of the resonance.

With these two quantities in hand, we have everything we need to quantize this normal mode.  The Hamiltonian (ignoring the zero-point energy) is given by the usual expression
\be
H=\hbar\Omega\, a^\dagger a,
\ee
and the physical observables $\hat \Phi$ and $\hat Q$ are given by Eq.~(\ref{eq:PHIQ}).  In the case of a general `black box' containing an arbitrary linear (purely reactive) circuit with a single input port, the admittance at that port will have a series of zeros corresponding to each internal collective mode as shown in Fig.~(fig:Admittance).

We should understand the raising and lowering operators as photon creation and destruction operators.  We are used to thinking of photons in free space, but here we are in a lumped element circuit.  The electric field of the photon mode lives between the capacitor plates and the magnetic field lives in a physically separate space inside the inductor.  We are also used to seeing photons introduced only in second-quantized notation.  However it can be instructive to use an ordinary first quantization approach to this simple harmonic oscillator.  Since we have chosen the flux $\Phi$ as the coordinate, we can easily write the ground state wave function for the oscillator as a function of this coordinate
\begin{eqnarray}
\Psi_0(\Phi) &=& \frac{1}{\sqrt{F}}e^{-\frac{\Phi^2}{4\Phi_\mathrm{ZPF}^2}}\\
\Psi_1(\Phi) &=& \frac{1}{\sqrt{F}}\frac{\Phi}{\Phi_\mathrm{ZPF}}e^{-\frac{\Phi^2}{4\Phi_\mathrm{ZPF}^2}}\\
\end{eqnarray}
where $F\equiv \sqrt{2\pi\Phi_\mathrm{ZPF}^2}$ is the normalization factor.

From this first-quantized representation we see that photon Fock states wave functions have definite reflection parity in $\Phi$ and hence Fock states (eigenstates of photon number) obey $\langle n|\hat\Phi|n\rangle=0$.   Only a coherent superposition of different Fock states can have a non-zero expectation value for the flux (or correspondingly, the electric field).  This is consistent with the second-quantized representation of $\hat \Phi$ in Eq.~(\ref{eq:hatPhi}) which is clearly off-diagonal in photon number. Inspired by the pioneering experiments of the Paris Rydberg atom cavity QED group\cite{HarocheRaimondRMP,Deleglise2008,HarocheRaimondcQEDBook} and using the methods of circuit QED, the Schoelkopf group first mapped a coherent superposition of qubit ground and excited states onto the corresponding coherent superposition of the zero- and one-photon states of a cavity and then by direct measurement of the electric field showed that the phase of the electric field matched the phase of the qubit superposition and that the average electric field vanished for Fock states\cite{Houck2007SinglePhoton}.  Following this achievement the Martinis group engineered remarkably complex arbitrary coherent superpositions of different photon Fock states, illustrating the level of quantum control that is enabled by the strong atom-photon coupling in circuit QED\cite{Hofheinz2008,Hofheinz2009}.  Very recently the Schoelkopf group taken advantage of the very strong dispersive coupling between qubit and cavity that can be achieved in circuit QED to generate large Schr\"odinger cat states for photons\cite{GerhardKerrCats}.

The generaliation of Eq.~(\ref{eq:PHIQ}) simply involves summing over all the normal modes
\begin{eqnarray}
\hat \Phi &=& \sum_j\Phi_\mathrm{ZPF}^{(j)} (a_j + a_j^\dagger)\label{eq:PHI2}\\
\hat Q &=& -i\sum_j Q_\mathrm{ZPF}^{(j)} (a_j - a_j^\dagger)
\label{eq:PHIQ2}
\end{eqnarray}
where
\begin{eqnarray}
\Phi_\mathrm{ZPF}^{(j)} &\equiv& \sqrt{\frac{\hbar Z_j}{2}}\\
Q_\mathrm{ZPF}^{(j)} &\equiv& \sqrt{\frac{\hbar}{2Z_j}}
\end{eqnarray}
involves the impedance $Z_j$ of the $j$th mode as viewed from the input port.

\subsection{Superconducting Qubits}
In order to go beyond the simple $LC$ harmonic oscillator to create a qubit, we need a non-linear element to produce anharmonicity in the spectrum.  The non-linear circuit element of choice in superconducting systems is the Josephson tunnel junction.
%
%There are a number of different circuit designs and topologies which yield qubits of different types: flux, phase, and charge.  For simplicity, we will focus here on the charge-based `transmon' qubit which consists of a small dipole antenna whose two halves are connected by a Josephson junction as shown in Fig.~({fig:transmon}).
%
The first evidence that Josephson tunneling causes the Cooper pair box to exhibit coherent superpositions of different charge states was obtained by Bouchiat \etal \shortcite{BouchiatCPBPhysScr1998}.  This was followed in 1999 by the pioneering experiment of the NEC group \shortcite{Nakamura1999} demonstrating time-domain control of the quantum state of the CPB using very rapid control pulses to modulate the offset charge.

The remarkable recent progress in creating superconducting quantum bits and manipulating their states has been summarized in several reviews \shortcite{Devoret2004,Esteve2005,Wendin_Shumeiko2006,Wendin_Shumeiko2007,Clarke2008,WiringUpQuantumSystems,YouandNori2005,Nori2008,Korotkov2009}.  Nearly 30 years ago Leggett discussed the fundamental issues concerning the collective degrees of freedom in superconducting electrical circuits and the fact that they themselves can behave quantum mechanically \shortcite{Leggett1980}.  As noted earlier, the essential collective variable in a Josephson junction \shortcite{MartinisDevoretJJreview} is the phase difference of the superconducting order parameter across the junction.  The first experimental observation of the quantization of the energy levels of the phase `particle' was made by Martinis, Devoret and Clarke in 1985 \shortcite{MartinisDevoretClarke1985,Clarke1988Science}.

\begin{figure}[htpb]
\centering
\includegraphics[totalheight=2.5in,clip]{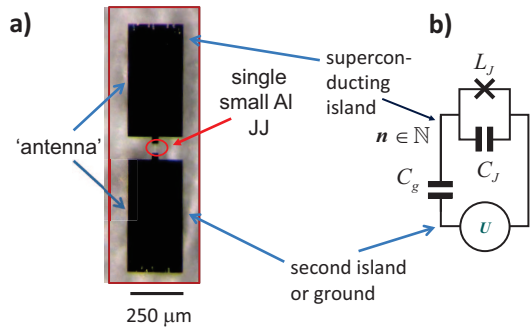}
\caption{a) Cooper pair box qubit (R. Schoelkopf lab) and b) its equivalent circuit showing a voltage source biasing the box through a coupling (`gate') capacitor $c_\mathrm{g}$. The cross denotes the Josephson junction. }
\label{fig:CPB}
\end{figure}

\begin{figure}[htpb]
\centering
\includegraphics[width=4.5in,clip]{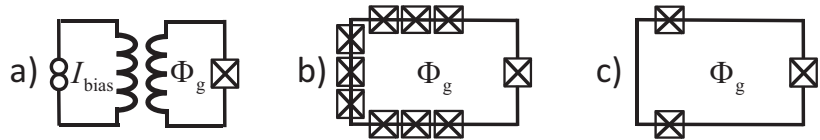}
\caption{Inductively shunted qubits. a) Phase qubit with a transformer flux bias circuit driven by current $I_\mathrm{bias}$.  Josephson junction is indicated by box with cross. b) Fluxonium qubit.  The shunt inductor has been replaced by an array of a large number of Josephson junctions.  The array junctions are chosen to have a sufficiently large ratio of Josephson energy $E_\mathrm{J}$ to charging energy $E_\mathrm{C}$ that phase slips can be neglected and the array is a good approximation to a very large inductor.  Flux bias circuit not shown. c) Flux qubit consisting of a superconducting loop with three Josephson junctions.  Flux bias circuit not shown.}
\label{fig:inductivelyshuntedqubits}
\end{figure}

\begin{figure}[htpb]
\centering
\includegraphics[totalheight=2.5in,clip]{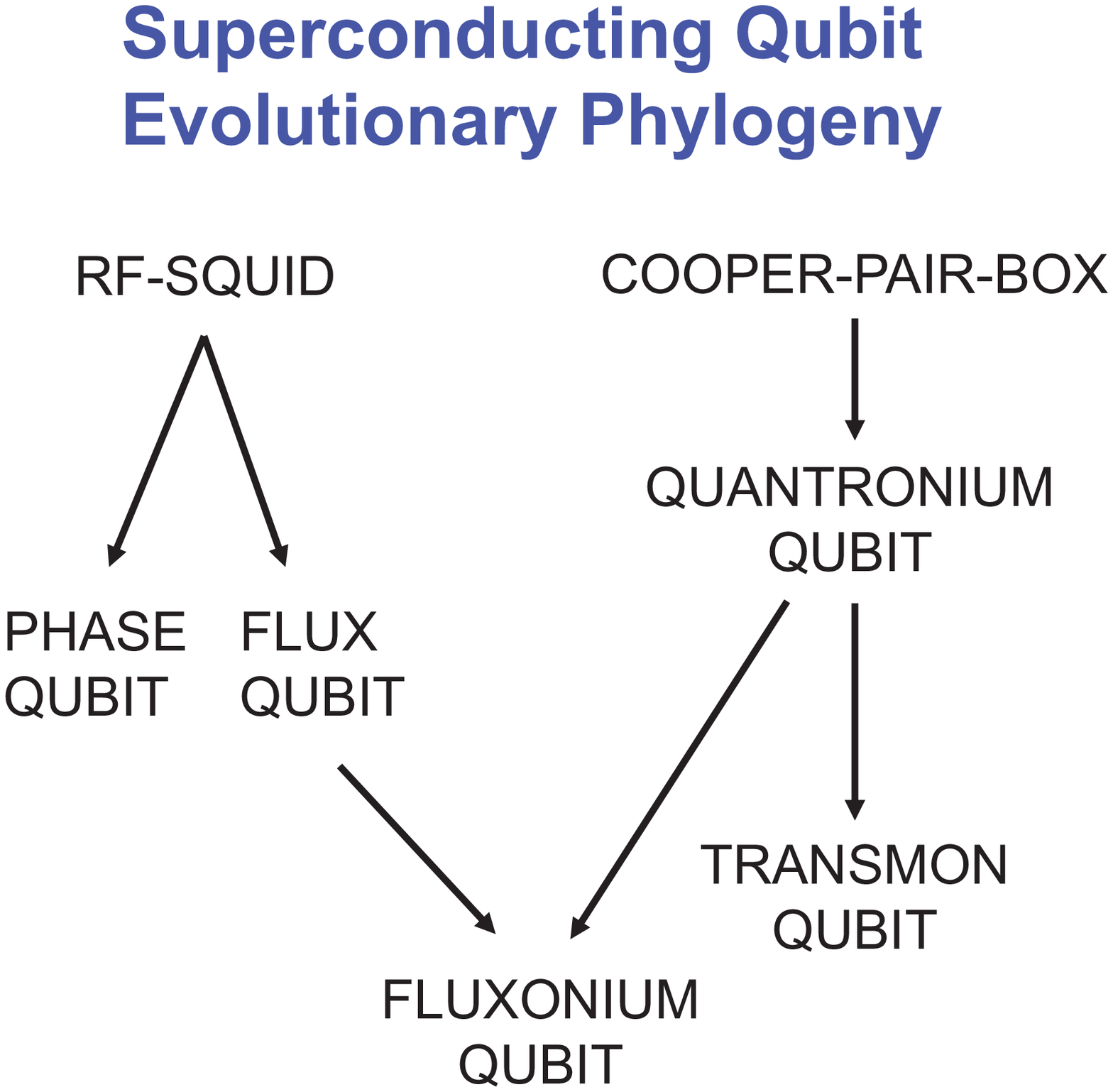}
\caption{Evolutionary phylogeny of superconducting qubits. (Courtesy M. Devoret.)}
\label{fig:qubitphylogeny}
\end{figure}

\begin{figure}[htpb]
\centering
\includegraphics[width=4.5in,clip]{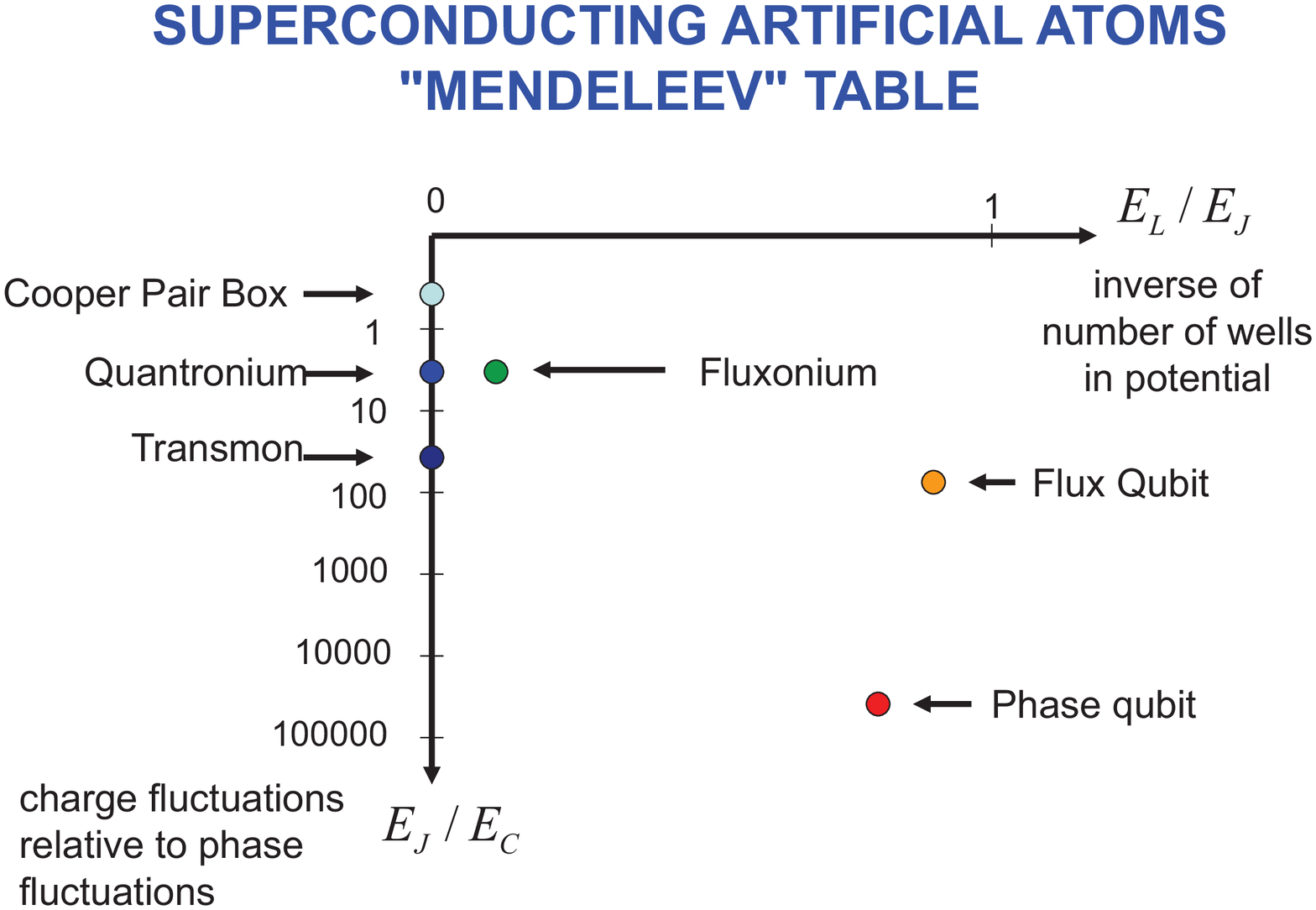}
\caption{`Periodic Table' of superconducting qubits. $\EJ$ is the tunneling Josephson energy, $4\EC$ is the energy cost to charge the junction with one Cooper pair, and $\EL/2$ is the energy cost to `charge' the shunt inductor with one flux quantum. (Courtesy M. Devoret.)}
\label{fig:MendeleevTable}
\end{figure}

A number of different qubit designs, illustrated in Fig.~(\ref{fig:CPB}) and Fig.~(\ref{fig:inductivelyshuntedqubits}) have been developed around the Josephson junction including the Cooper pair box (CPB) \shortcite{Averin1985,Buttiker1987,Lafarge1993,BouchiatCPBPhysScr1998,Nakamura1999,Vion2002,Koch2007,Schreier2008} based on charge, the flux qubit \shortcite{Mooij1999,Wal2000,Chiorescu2003}, and the phase qubit \shortcite{Martinis2002,Berkley2003}.  Devoret and co-workers have recently introduced the fluxonium qubit \shortcite{fluxonium,Manucharyan_coherent_oscillations} in which the small Josephson junction is shunted by a very high inductance created from a string of larger Josephson junctions.  Fig.~(\ref{fig:qubitphylogeny}) shows an `evolutionary phylogeny' for these different types of qubits and Fig.~(\ref{fig:MendeleevTable}) classifies them into a `periodic table' of the `elements' according to the relative size of the Coulomb charging energy, the Josephson energy and the energy stored in the shunt inductor.

  We will will not review here the Hamiltonians of these different types of qubits and their relative merits in terms of their sensitivity to noise perturbations but instead focus on the Cooper pair and the so-called `transmon qubit' which holds the current records for phase coherence time.

\subsection{The Cooper Pair Box and the Transmon Qubit}

The Cooper pair box (CPB) \shortcite{MichelLesHouches} is topologically distinct from the other designs shown in Fig.~(\ref{fig:inductivelyshuntedqubits}) in that it has no wire closing the loop around the junction.  It consists very simply of a small antenna whose two halves are connected by a Josephson junction.
 The Hamiltonian will be described below.  Because the two sides of the junction are not shunted by an inductor, the number of Cooper pairs transferred through the junction is a well-defined integer.  The integer charge implies the conjugate phase is compact; that is, in the phase representation, the system obeys periodic boundary conditions.  This implies that charge-based qubits are sensitive to stray electric field noise, but that this can be overcome by putting the Cooper pair box in the `transmon' regime where the Josephson tunneling energy dominates over the Coulomb charging energy \shortcite{Koch2007,LifeAfterChargeNoise2009}.

 The Lagrangian for the Cooper pair box qubit is similar to that of the $LC$ oscillator studied above except that unlike the inductor, the energy stored in the Josephson junction is not quadratic in the flux but rather periodic
 \be
 L=\frac{1}{2}C_\Sigma \dot\Phi^2 + \EJ \cos\varphi,
 \ee
 where $C_\Sigma=C_\mathrm{geom} + C_\mathrm{J}$ is the sum of the geometric capacitance of the antenna plus the Josephson tunnel junction capacitance, and $\varphi$ is the phase difference of the superconducting order parameter across the Josephson junction and is related to the usual flux variable by
 \be
 \varphi = 2\pi \frac{\Phi}{\Phi_0}= \frac{2e}{\hbar}\Phi,
 \ee
and $\Phi_0$ is the superconducting flux quantum.  Because there is no shunting inductor, the physical state of the Josephson junction is invariant under shifts of the phase of the order parameter by $2\pi$.  Hence $\varphi$ is an angular variable and the wave function obeys periodic boundary conditions.  It is convenient to work with $\varphi$ rather than $\Phi$ and so we write
\be
L=\frac{\hbar^2}{8e^2}C_\Sigma \dot\varphi^2 + \EJ \cos\varphi.
\ee
This is the Lagrangian of a quantum rotor in a gravitational field, where $\EJ$ plays the role of the strength of the gravitational field
and $C_\Sigma$ plays the role of the moment of inertia of the rotor.  The angular momentum conjugate to the angle $\varphi$ is
\be
n = \frac{\delta L}{\delta \dot\varphi} = \frac{\hbar^2}{4e^2}C_\Sigma \dot\varphi,
\ee
and the quantum Hamiltonian for the rotor may be written
\be
H = 4\EC [\hat n - n_g]^2 -\EJ\cos\hat\varphi,
\ee
and $\EC\equiv \frac{e^2}{2C_\Sigma}$ is the charging energy of a single electron (half a Cooper pair) stored on the total capacitance.  $\hat n \equiv -i\frac{\partial}{\partial\varphi}$ is the integer-valued angular momentum conjugate to the angle $\varphi$ and is the operator representing the integer number of Cooper pairs that have tunneled through the Josephson junction relative to the equilibrium classical charge state. Note that we have included in the Hamiltonian the so-called offset charge $n_g$ which represents external bias voltage as shown in Fig.~(\ref{fig:CPB}).  In addition to any intentionally applied external bias, $n_g$ can also contain random fluctuations due to stray charges jumping around in the underlying substrate or the Josephson junction barrier.  It turns out that in the limit $\EJ\gg\EC$ the low frequency noise in the offset charge can be neglected\shortcite{Koch2007,LifeAfterChargeNoise2009} resulting in very long coherence times for the transmon qubit and so we will henceforth neglect this term.

In the limit $\EJ\gg\EC$, the `gravitational' force is very strong and the `moment of inertia' is very large so the phase angle undergoes only very small quantum fluctuations around zero.  In this limit one can safely expand the cosine in a power series when studying the low-lying excitations\shortcite{Koch2007}
\begin{eqnarray}
H &=& H_0 + V,\\
H_0 &=& 4\EC\hat n^2 + \frac{\EJ}{2}\hat\varphi^2,\\
V &=& \EJ[-\frac{1}{4!}\hat\varphi^4 + \frac{1}{6!}\hat\varphi^6+\ldots]
\end{eqnarray}
We see that $H_0$ is the Hamiltonian of a simple $LC$ harmonic oscillator with frequency $\hbar\Omega = \sqrt{8\EJ\EC}$ and with the leading order effect of the Josephson junction being to play the role of a (relatively large) linear inductor whose energy is quadratic in the phase (flux) across it.  In the limit $\EJ\gg\EC$ we are effectively ignoring the fact that $\varphi$ is an angular variable and taking the harmonic oscillator coordinate $\varphi$ to be non-compact and imposing vanishing boundary conditions at infinity on the wave functions.  It is this assumption which makes conjugate momentum $\hat n$ continuous rather than integer-valued and also allows us to neglect the offset charge term.\cite{Koch2007}  Essentially the large charge fluctuations associated with the small phase fluctuations wash out the discreteness of the charge.

In terms of second quantization, we have
\be
\hat\varphi=\varphi_\mathrm{ZPF}[b+b^\dagger]
\ee
with $\varphi_\mathrm{ZPF}=\frac{\hbar\Omega}{2\EJ}$.  The Hamiltonian becomes
\begin{eqnarray}
H_0 &=& \hbar\Omega b^\dagger b\\
V &=& -\frac{\alpha}{2}(b+b^\dagger)^4 + \ldots \approx -\alpha b^\dagger b -\frac{\alpha}{2} b^\dagger b^\dagger bb
\end{eqnarray}
where we have made the rotating wave approximation in the last term of $V$. The anharmonicity is given by\cite{Koch2007}
\be
\alpha\approx \EC
\ee
and the first term in $V$ leads to a small renormalization of the bare transition frequency
\be
\hbar\tilde\Omega \approx \hbar\Omega - \EC
\label{eq:freqrenorm}
\ee

The perturbation term $V$ represents the fact that the Josephson energy is not simply quadratic in flux as in a linear inductor.  We are effectively dealing with a non-linear inductor which causes the system to have the anharmonic spectrum we need to use it as a qubit.  The transition frequency from ground to first excited state is $\hbar\Omega_{01}=\hbar\tilde\Omega$ while the next transition in the ladder is $\hbar\Omega_{12} = \hbar\Omega_{01}-\EC$.  In a typical transmon qubit\shortcite{Koch2007,LifeAfterChargeNoise2009} with $\EJ/\EC\sim 50-100$, this corresponds to a $\sim 3-5\%$ negative anharmonicity of $\sim 200 \mathrm{MHz}$.  This anharmonicity is large enough that smooth microwave pulses with durations on the scale of a few nanoseconds will be sufficiently frequency selective that leakage into higher levels of the transmon can be neglected and we can treat it as a two-level qubit.

\subsection{Circuit QED: Qubits Coupled to Resonators}

We have seen that the presence of the Josephson junction causes the transmon to become a weakly anharmonic oscillator.  If we couple this qubit to a lumped element $LC$ resonator, a coplanar waveguide resonator or a 3D cavity, we will have a system analogous to that studied in cavity QED.  In this `circuit QED' setup the coupling to the microwave mode(s) of the resonator can be very strong.  Focusing on the case where only a single mode of the resonator is important and assuming that the qubit does not strongly perturb the resonator mode, our Hamiltonian becomes
\be
H = \hbar\tilde\Omega b^\dagger b -\frac{\alpha}{2} b^\dagger b^\dagger b b + \hbar\omega_\mathrm{r} a^\dagger a + \hbar g [a^\dagger b + a b^\dagger],
\ee
where $g$ represents the transition dipole coupling strength between the qubit and the resonator.   We have noted that the transmon anharmonicity is weak.  If despite this, the coupling to the cavity mode is sufficiently weak relative to the anharmonicity, $g\ll\alpha$, then we can limit our Hilbert space to the ground and first excited states of the qubit and treat it as a two-level system by making the substitutions
\begin{eqnarray}
b^\dagger b &\longrightarrow& \frac{1+\sigma^z}{2}\\
b &\longrightarrow& \sigma^-\\
b^\dagger &\longrightarrow& \sigma^+
\end{eqnarray}
which leads us (up to irrelevant constants and making a rotating wave approximation) to the standard Jaynes-Cummings Hamiltonian of cavity QED\cite{HarocheRaimondcQEDBook}
\be
H = \hbar\omega_\mathrm{r} a^\dagger a + \frac{\hbar\tilde\Omega}{2}\sigma^z +\hbar g [a\sigma^+ + a^\dagger\sigma^-].
\ee

In the dispersive limit where the qubit is detuned by a distance $\Delta\gg g$ from the cavity, applying a unitary transformation which diagonalizes the Hamiltonian to lowest order in $g$ yields the dispersive Hamiltonian\cite{BlaisCQEDtheory2004,HarocheRaimondcQEDBook}
\be
H = \hbar\omega_\mathrm{r} a^\dagger a + \frac{\hbar\tilde\Omega}{2}\sigma^z +\hbar \chi a^\dagger a\sigma^z.
\ee
where the dispersive coupling is given by $\chi=-g^2/\Delta$.  We see that the dispersive coupling causes the frequency of the qubit to depend on the photon occupation of the cavity \cite{Schuster2007,BlakeQND2010} and the frequency of the cavity to depend on the state of the qubit\cite{BlaisCQEDtheory2004}.  Thus this term permits QND readout of the state of the qubit by measuring the cavity frequency (e.g. by measuring the phase shift of photons reflected from the cavity) or QND readout of the state of cavity by measurement of the transition frequency of the qubit\cite{BlakeQND2010}.

In practice the coupling $g\sim 100-200 \mathrm{MHz}$ is typically sufficiently strong and the qubit anharmonicity is sufficiently small that the above derivation (in which the two-level qubit approximation is made first) is not quantitatively accurate. In addition the approximation that the cavity mode is not distorted by the presence of the qubit is often inaccuracte.  Nigg et al.\cite{NiggBBQ} have developed a much more careful treatment which does not treat the qubit cavity coupling perturbatively and explicitly takes advantage of the weak anharmonicity of the qubit(s).  This so-called `Black Box Quantization' (BBQ) scheme uses a commercial large-scale numerical finite-element Maxwell equation solver to (`exactly') solve for the normal modes of the linear system consisting of the cavity strongly coupled to the qubit (treated as a simple harmonic oscillator by replacing the Josephson cosine term by its leading quadratic approximation).  The normal modes computed this way can be arbitrarily distorted by strong coupling between the (multiple) cavity modes and the qubit and thus form a highly efficient basis in which to express the weak anharmonic term in the Josephson Hamiltonian.   The harmonic Hamiltonian for the system treats the qubit and the cavity modes all on an equal footing and is given by
\be
H_0 = \sum_j \hbar\omega_j a^\dagger_j a_j,
\ee
where the frequencies of the normal modes are determined by the locations of the numerically computed of the zeros of the admittance $Y(\omega)$ at a port across the Josephson inductance.

The Josephson phase variable $\varphi$ (or equivalently the flux at a network port defined across the Josephson inductance) can (by analogy with Eq.~(\ref{eq:PHI2})) be precisely expressed in terms of the $j$th normal mode operators as
\be
\hat\varphi = \sum_j \varphi_\mathrm{ZPF}^{(j)} [a^\dagger_j + a_j].
\ee
Because the quadratic term in the expansion of the cosine has already been included in the harmonic part of the Hamiltonian, the next leading term is
\be
V=-\frac{\EJ}{4!}\hat\varphi^4 = -\frac{\EJ}{4!}\left\{\sum_j \varphi_\mathrm{ZPF}^{(j)} [a^\dagger_j + a_j]\right\}^4.
\ee
Normal ordering and making rotating wave approximations (RWA) leads to corrections to normal mode frequencies analogous to Eq.~(\ref{eq:freqrenorm}).  In addition every mode develops some anharmonicity (self-Kerr) inherited from the Josephson junction.  The most anharmonic mode can be identified as `the qubit' and the remainder as `cavity modes.'  In addition to self-Kerr terms, there will be cross-Kerr terms in which the frequency of a given mode depends on the occupation numbers of other modes.  The general form of the non-linear Hamiltonian within the RWA is
\be
V = \sum_j \delta\omega_j \hat n_j + \frac{1}{2}\sum_{j,k}\chi_{jk} \hat n_j \hat n_k
\ee
where $\hat n_j = a^\dagger_j a_j$ is the occupation number operator for the $j$th mode.  At this point it is now generally safe and accurate to map the most anharmonic component (aka the qubit) onto a spin-1/2 as done above.  The procedure outlined here is much more accurate than making the two-level approximation first and treating the coupling to the cavity perturbatively, especially in the case where the anharmonicity is weak and the coupling $g$ is strong.  It should not be forgotten that there may be contributions from the sixth and higher-order terms in the expansion of the cosine potential which have been neglected here.

One of the remarkable aspects of circuit QED is that it is easy to reach the strong-dispersive limit where some elements of the $\chi$ matrix are $\sim 10^1 - 10^3$ times larger than the line widths of both the qubit and the cavity.  This makes it possible for example to create photon cat states using the self-Kerr non-linearity which the cavity inherits from the qubit\cite{GerhardKerrCats}.  This strong dispersive coupling also opens up a new toolbox for mapping qubit states onto cavity cat states\cite{ZakiQCMAP} and possibly even for autonomous error correction using cavity states as memories\cite{ZakiAutoQEC}.   It is also worth noting from the form of the cross-Kerr Hamiltonian that both qubits and cavities can be readily dephased by stray photons hopping into and out of higher cavity modes that would otherwise be considered irrelevant.  Hence very careful filtering to remove these stray photons is important.

\section{Summary}

These notes have attempted to convey some of the basic concepts in quantum information as well as provide a basic introduction to circuit QED. A much more detailed review of circuit QED is presented in the author's lecture notes in the Proceedings of the 2011 Les Houches Summer School on Quantum Machines\cite{LesHouchesQuantumMachines}.

\section{Acknowledgments}

The author is grateful for helpful conversations with numerous colleagues including among others Charles Bennett, Liang Jiang, Sreraman Muralidharan, Uri Vool, Claudia De Grandi, Simon Nigg, Matt Reed, Michael Hatridge, Shyam Shankar, Gerhard Kirchmair, Robert Schoelkopf and Michel Devoret.  The author's research is support by Yale University, the NSF, the ARO/LPS and IARPA.

This manuscript will appear as a chapter in the book `LECTURE NOTES ON STRONG LIGHT-MATTER COUPLING: FROM ATOMS TO SOLID STATE SYSTEMS,' published by World Scientific, Singapore.

%%%%%%%%%%%%%%%%%%%%%%%%%%%%%%

\bibliographystyle{ws-rv-van}
%\bibliographystyle{ws-rv-har}
%\bibliography{ws-rv-sample}

\bibliography{Girvin_LesHouchesFromSolvay_Bibv4_Repaired}

\begin{thebibliography}{106}
\bibitem{jointmeasurement}
Alice can perform a unitary transformation which un-entangles the four Bell
  states, mapping them into four distinct states in the measurement basis
  ($|00\rangle, |01\rangle, |10\rangle, |11\rangle$). Then simply measuring
  both qubits tells her which of the four Bell states she had received.
\end{thebibliography}

\end{document}